\documentclass[preprint]{aastex} 

\newif\ifimage
\imagetrue

\newif\ifthesis
\thesisfalse

\usepackage{graphicx}
\usepackage{natbib}
\usepackage{afterpage}
\usepackage{lscape}
\usepackage{longtable}
\usepackage{placeins}

\defcitealias{Ianj2012}{I2012}
\defcitealias{StilpGlobal}{Paper~I}
\defcitealias{StilpResolved}{Paper~II}

\newcommand{\rs}{\ensuremath{r_s}}
\newcommand{\ps}{\ensuremath{p_s}}

\newcommand{\fcold}{\ensuremath{f_\mathrm{cold}}}

\newcommand{\um}{$\mu$m}
\newcommand{\hi}{\mbox{\rm H\,{\sc i}}}

\newcommand{\m}[1]{\ensuremath{M_\mathrm{#1}}}
\newcommand{\msun}{\ensuremath{\mathrm{M}_{\odot}}}

\newcommand{ \kms}{km~s$^{-1}$}

\newcommand{\halpha}{\mbox{\rm H$\alpha$}}
\newcommand{\degrees}{$^{\circ}$}
\newcommand{\gh}{Gauss-Hermite}
\newcommand{\mhi}{\ensuremath{M_{\mathrm{HI}}}}
\newcommand{\sfrsd}{\ensuremath{\Sigma_{\mathrm{SFR}}}}

\newcommand{\sd}[1]{\ensuremath{\Sigma_{[1]}}}

\newcommand{\per}[2][1]{#2$^{-#1}$}
\newcommand{\sumlim}[2][]{\displaystyle\sum\limits_{#2}^{#1}}

\newcommand{\citeeg}[1]{\citep[e.g.,][]{#1}}

\newcommand{\vp}{\ensuremath{v_{\mathrm{peak}}}}

\newcommand{\scentral }{\ensuremath{\sigma_{\mathrm{central}}}}

\newcommand{\swing}{\ensuremath{ \sigma_{\mathrm{wings}}}}
\newcommand{\swingsq}{\ensuremath{ \sigma^2_{\mathrm{wings}}}}
\newcommand{\fw}{\ensuremath{f_{\mathrm{wings}}}}

\newcommand{\mstarcumul}[1]{\ensuremath{\m{\star}(< #1)}}

\newcommand{\smomtwo}{\ensuremath{\sigma_\mathrm{m2}}}

\newcommand{\avesfr}[1]{\ensuremath{\mathrm{SFR}(#1)}}

\newcommand{\avesfrsd}[1]{\ensuremath{\Sigma_\mathrm{SFR}(#1)}}

\newcommand{\figurepath}{./figures}
\newcommand{\tablepath}{./tables}

\shorttitle{Timescales on which Star Formation Couples to the Neutral ISM}

\begin{document}

\title{Timescales on which Star Formation Affects the Neutral ISM}

\author{
Adrienne M. Stilp\altaffilmark{1},
Julianne J. Dalcanton\altaffilmark{1},
Steven R. Warren\altaffilmark{2},
Daniel R. Weisz\altaffilmark{1},
Evan Skillman\altaffilmark{3},
J\"{u}rgen Ott\altaffilmark{4},
Benjamin F. Williams\altaffilmark{1},
Andrew E. Dolphin\altaffilmark{5}
%B\"{a}rbel Koribalski\altaffilmark{5}
}

\altaffiltext{1}{Department of Astronomy, University of Washington, Box 351580, Seattle, WA 98195,
USA}

\altaffiltext{2}{Department of Astronomy, University of Maryland, CSS Bldg., Rm. 1024, Stadium Dr., College Park, MD 20742-2421}

\altaffiltext{3}{Minnesota Institute for Astrophysics, University of Minnesota, 116 Church St. SE,
Minneapolis, MN 55455, USA}

\altaffiltext{4}{National Radio Astronomy Observatory, P.O. Box O, 1003
  Lopezville Road, Socorro, NM 87801, USA}

%\altaffiltext{5}{Australia Telescope National Facility, CSIRO Astronomy and Space Science, PO Box 76,
%Epping NSW 1710, Australia}

\altaffiltext{5}{Raytheon Company, 1151 East Hermans Road, Tucson, AZ 85756, USA }

\ifthesis
\chapter{Timescales on which SF Couples to the Neutral ISM}
\label{angst}
\fi

\ifthesis
\renewcommand{\figurepath}{./chapters/angst/figures}
\renewcommand{\tablepath}{./chapters/angst/tables}
\else
\renewcommand{\figurepath}{.}
\renewcommand{\tablepath}{.}
\fi

\ifthesis
\else
\begin{abstract}
\fi

Turbulent neutral hydrogen (\hi{}) line widths are often thought to be driven primarily by star formation (SF), but the timescale for converting SF energy to \hi{} kinetic energy is unclear.
As a complication, studies on the connection between \hi{} line widths and SF in external galaxies often use broadband tracers for the SF rate, which must implicitly assume that SF histories (SFHs) have been constant over the timescale of the tracer.
In this paper, we compare measures of \hi{} energy to time-resolved SFHs in a number of nearby dwarf galaxies.
We find that \hi{} energy surface density is strongly correlated only with SF that occurred $30 - 40$~Myr ago.
This timescale corresponds to the approximate lifetime of the lowest mass supernova progenitors ($\sim 8$ \msun{}).
This analysis suggests that the coupling between SF and the neutral ISM is strongest on this timescale, due either to an intrinsic delay between the release of the peak energy from SF or to the coherent effects of many SNe during this interval.
At $\sfrsd{} > 10^{-3}$ \msun{} \per{yr} \per[2]{kpc}, we find a mean coupling efficiency between SF energy and \hi{} energy of $\epsilon = 0.11 \pm 0.04$ using the $30 - 40$~Myr timescale.
However, unphysical efficiencies are required in lower \sfrsd{} systems, implying that SF is not the primary driver of \hi{} kinematics at $\sfrsd{} < 10^{-3}$ \msun{} \per{yr} \per[2]{kpc}.

\ifthesis
\else
\end{abstract}

\fi

\section{Introduction}
\label{angst::sec:intro}

Neutral hydrogen (\hi{}) line widths in galaxies are commonly thought to be due to turbulence driven by star formation (SF), which releases energy through ionizing radiation, stellar winds, and supernova explosions \citep[SNe; e.g.,][]{Spitzer1978, MacLow2004}.
Because turbulence decays rapidly in typical ISM conditions \citep[$\tau \sim 10$~Myr;][]{MacLow1999}, any driver must continuously replenish the observed turbulent energy.
At high SFR surface density (\sfrsd{}), a number of studies have found strong correlations between the amount of SF and \hi{} turbulence, and have used these correlations to constrain possible mechanisms by which SF couples to the neutral ISM \citeeg{Tamburro2009, Joung2009}.
At low \sfrsd{}, however, there appears to be little connection between SF intensity and \hi{} velocity dispersion as found in dwarf galaxies or the outer disks of spirals \citep{vanZee1999, Dib2006, Tamburro2009}.
The primary driver of turbulence remains unknown in these low \sfrsd{} regimes.

%In addition to uncertainties in the driving mechanisms of turbulence, the timescale for driving \hi{} turbulence is also unknown.
In addition to uncertainties in the driving mechanism of turbulence, there are few constraints on the timescale over which the energy that drives turbulence is injected into the ISM.
This timescale cannot be calculated from first principles, and is challenging to constrain observationally because the timescale for energy input from SF is also uncertain.
Common tracers of the SF rate (SFR) are either \halpha{} or far ultraviolet (FUV) emission, occasionally with the inclusion of far-infrared (FIR) fluxes to provide an estimate of dust-obscured SF \citeeg{Leroy2008, Leroy2012, Kennicutt2012}.
FUV wavelengths trace emission from SF over the past $10-100$ Myr,
 while \halpha{} typically probes much shorter timescales of $3-10$~Myr \citep[][and references therein]{Kennicutt2012}.
The calibration of these tracers relies on the assumption of a constant SFH over the past $\sim 10 - 100$~Myr, and is not yet well-understood for low-SFR systems with large relative variations.

In contrast, time-variable SFHs in galaxies have been well-established \citeeg{Grebel1997, Mateo1998, Dolphin2005, Weisz2008, Weisz2011}, which complicates the interpretation of FUV- or \halpha{}-based SFR indicators, but time-resolved SFHs are difficult to obtain for many galaxies.
Additionally, it has recently been shown that the SFR traced by FUV or \halpha{} emission may not be well-matched to the actual time-resolved SFH \citeeg{McQuinn2010, Johnson2012}.
These SFR estimates also do not allow for a measurement of the impact of SF over time on the surrounding ISM.
Moreover, in dwarf galaxies, there are discrepancies between SFRs as traced by FUV emission and by \halpha{} \citeeg{Lee2009, Meurer2009, Boselli2009}.
Potential solutions to this problem are stochastic sampling of both the IMF and the cluster mass function, variable SFRs, a variable IMF, or a combination of these possibilities \citeeg{Fumagalli2011, Lee2011, Weisz2012, daSilva2012}.
More robust comparisons between time-resolved recent SFHs and ISM properties are necessary to constrain the timescales for energy input from SF and the response of the ISM.

In this paper, we address the issue of timescales for energy input from SF using the combination of two unique data sets.
The first is the ACS Nearby Galaxy Survey Treasury program \citep[ANGST; ][]{Dalcanton2009}, which observed 69 galaxies within $\sim 4$ Mpc.
These data have been used to produce time-resolved SFHs from galaxies' resolved stellar populations \citeeg{Williams2009, Gogarten2010, Weisz2011, Williams2011}.
The second dataset is composed of high resolution \hi{} observations from the follow-up Very Large Array-ANGST project \citep[``VLA-ANGST'';][]{Ott2012} and The \hi{} Nearby Galaxy Survey \citep[``THINGS'';][]{Walter2008}.
Our sample is composed primarily of dwarf galaxies.
These systems have both lower gravitational wells compared to spirals, which should enhance the effects of energy deposition from SF, as well as larger scale heights, which should more easily contain the energy released from SF within the disk.
With these two datasets, we search for a preferred timescale over which energy input from SF transfers to the surrounding neutral ISM.
In \S~\ref{angst::sec:sample}, we discuss the sample selection and the data used to address this question.
In \S~\ref{angst::sec:analysis}, we calculate \hi{} energies and \sfrsd{} values.
In \S~\ref{angst::sec:comparisons} we assess the level of correlation between \hi{} energy and SF on a variety of timescales.
In \S~\ref{angst::sec:discussion}, we discuss the implied coupling efficiencies between SF energy and \hi{} energy, as well as the physical causes that drive the observed correlations.
Finally, in \S~\ref{angst::sec:summary}, we summarize our results.

%
% Data
%

\section{Sample and Data}
\label{angst::sec:sample}

We first introduce the general properties of our sample in \S~\ref{angst::sec:sample--sample}.
We then briefly describe the data used to derive SFHs and \hi{} properties in \S~\ref{angst::sec:sample--sfhs} and \ref{angst::sec:sample--hi}.

\subsection{The Sample}
\label{angst::sec:sample--sample}

Our sample consists of a subset of ANGST galaxies \citep{Dalcanton2009} that have high-quality \hi{} observations through either VLA-ANGST or THINGS.
In
\ifthesis
\citetalias{StilpGlobal},
\else
\citet[][hereafter Paper I]{StilpGlobal},
\fi
 we presented analysis of \hi{} kinematics on global scales by co-adding individual \hi{} line-of-sight spectra after removal of the rotational velocity.
We select the sample for this paper from that in \citetalias{StilpGlobal}.
The original selection criteria are described in detail in \citetalias{StilpGlobal}, but we briefly review them here:
\begin{enumerate}

\item \hi{} instrumental spatial resolution smaller than our working resolution of 200~pc, which is the approximate spatial resolution at the limit of the ANGST survey;

\item Velocity resolution $\Delta v \leq 2.6$ \kms{}, to accurately determine the peak of each \hi{} line-of-sight spectrum and its corresponding velocity (\vp{}), as well as the average line width;

\item Inclination $< 70$\degrees{}, as line-of-sight profiles for galaxies with larger inclinations may be artificially broadened;

\item No noticeable contamination from the Milky Way or a companion, to accurately determine \vp{};

\item More than 10 independent beams across the galaxy disk above the signal-to-noise threshold ($S/N > 5$).

\end{enumerate}
These choices maximize our data quality, as described more fully in \citet{StilpGlobal}.

Applying these criteria leave us with 18 galaxies, primarily with de Vaucouleurs T-type of 10 (i.e., dwarf irregulars), as many of the more massive spirals were eliminated based on the first or second criteria.
General properties of the sample are given in Table~\ref{tab:angst--sample}.
Galaxies are listed in decreasing total baryonic mass (\m{baryon,tot}).
We give
(1) the galaxy name;
(2) the survey for \hi{} data;
(3-4) position in J2000 coordinates;
(5) distance in Mpc from \citet{Dalcanton2009};
(6) inclination from \citetalias{StilpGlobal}, with the exception of Sextans~B where an $i \sim 30$\degrees{} better matches the \hi{} morphology compared with the inclination quoted in \citetalias{StilpGlobal};
(7) \m{baryon,tot} from \citetalias{StilpGlobal};
(8) total \hi{} mass \mhi{} from \citetalias{StilpGlobal};
(9) average SFR in the ANGST aperture over the past 100~Myr derived from ANGST SFHs (see \S~\ref{angst::sec:sample--sfhs});
(10) de Vaucouleurs T-type.
% XXX add SFR from FUV+24\um{} to this table.

We show the inclination-corrected \hi{} column density maps of our sample in Figure~\ref{angst::fig:footprints}, with the ANGST footprints overlaid.
While a few of the galaxies in the sample have disk-like morphologies, the majority are dwarf irregulars.
In some cases, the ANGST aperture is fully covered by \hi{} with $S/N > 5$.
In others, the $S/N > 5$ region is smaller than the ANGST aperture, but in these galaxies the majority of the SF is located in the same region as the high $S/N$ \hi{}.

\begin{figure}
\centering
\includegraphics{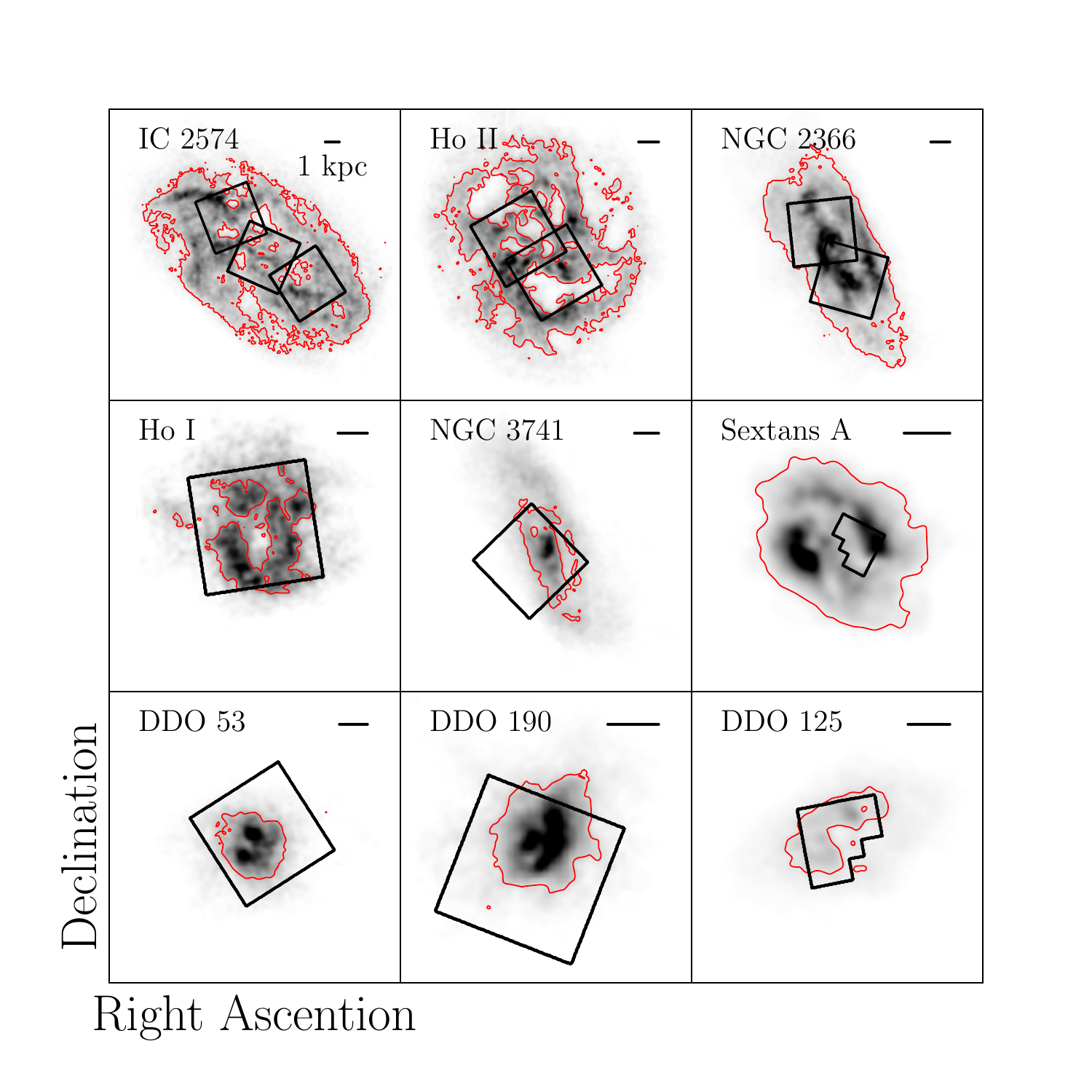}
\caption{ \hi{} column density maps for the sample, scaled between 0 and $2 \times 10^{21}$ cm$^{-2}$.
The ANGST footprints are overlaid as the thick black solid line.
We also show the $S/N = 5$ level of the \hi{} spectra as the red contour.
Only regions that are both inside the ANGST footprint and with $S/N > 5$ are included in our analysis.
The scale bar indicates 1 kpc in all panels, and the physical resolution of all \hi{} data is 200 pc at the distance of each galaxy.
\label{angst::fig:footprints}}
\end{figure}
\addtocounter{figure}{-1}
\begin{figure}
\centering
\includegraphics{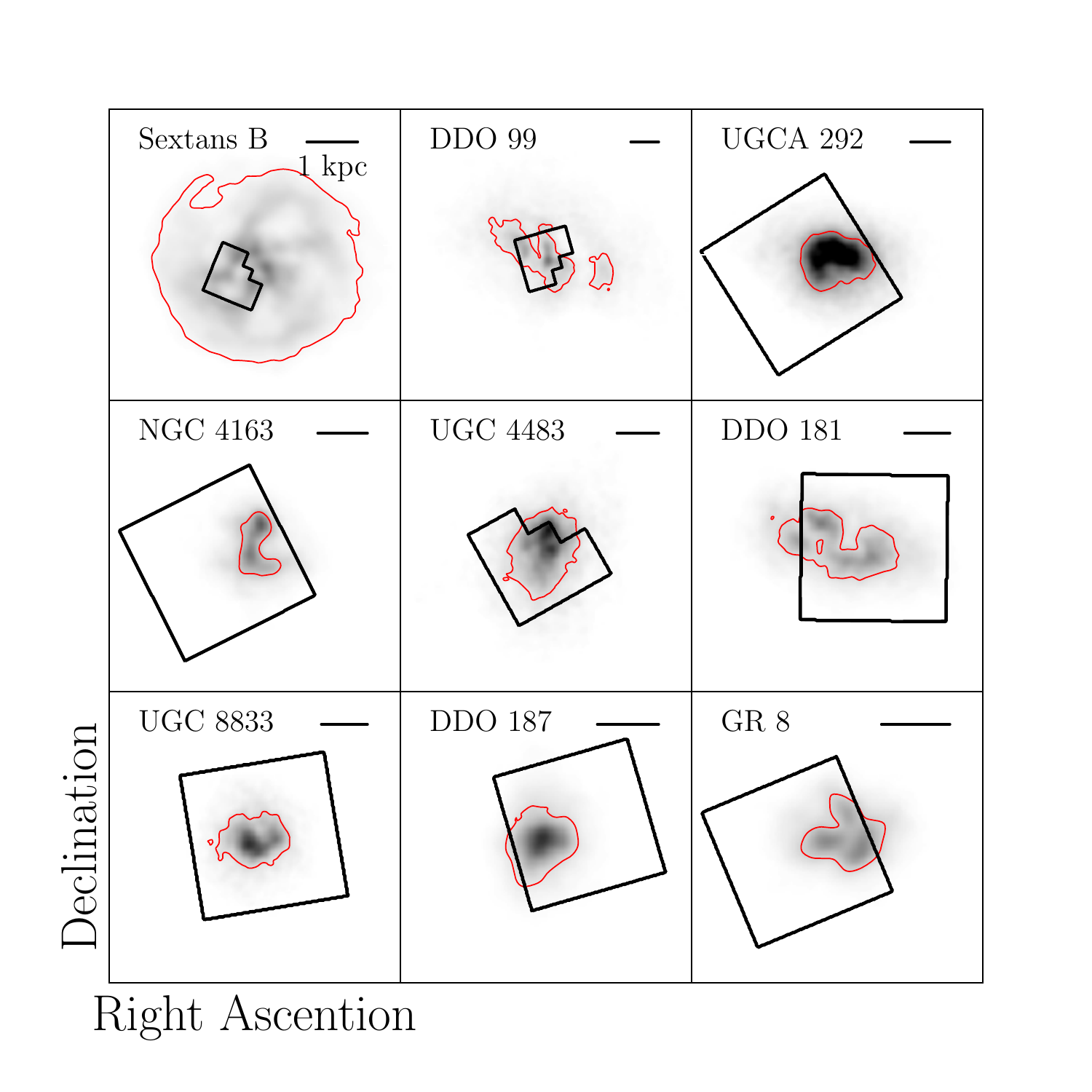}
\caption{ continued.}
\end{figure}

\ifthesis
\afterpage{\input{\tablepath/table-sample.tex}}
\else
\afterpage{\begin{deluxetable}{lccccccccc}
\tabletypesize{\scriptsize}
\rotate
\tablewidth{0pt}
\tablecaption{ANGST \hi{} kinematics sample. \label{tab:angst--sample}}
\tablehead{

  \colhead{1} &
  \colhead{2} &
  \colhead{3} &
  \colhead{4} &
  \colhead{5} &
  \colhead{6} &
  \colhead{7} &
  \colhead{8} &
  \colhead{9} &
  \colhead{10} \\

  \colhead{Galaxy} & 
  \colhead{Survey} & 
  \colhead{RA} & 
  \colhead{Dec} & 
  \colhead{Distance} & 
  \colhead{$i$} & 
  \colhead{\m{baryon,tot}} & 
  \colhead{\m{HI,tot}}  &  
  \colhead{$\langle \mathrm{SFR} (< 100 \; \mathrm{Myr}) \rangle$} & 
  \colhead{Type} \\
  
  \colhead{} &  
  \colhead{} & 
  \colhead{(hh:mm:ss)} & 
  \colhead{(dd:mm:ss)} & 
  \colhead{(Mpc)} & 
  \colhead{(\degrees{})} & 
  \colhead{(log \msun{})} &
  \colhead{(log \msun{})} & 
  \colhead{(10$^{-3}$ \msun yr$^{-1}$)} &
  \colhead{}   \\

}

\startdata
IC 2574  &  THINGS  &  10:28:27.7  &  +68:24:59  &  3.79  &  55  &  9.4  &  9.1  &  82.5  &   9  \\ 
Ho II  &  THINGS  &  08:19:05.0  &  +70:43:12  &  3.38  &  49  &  9.0  &  8.8  &  67.1  &  10  \\ 
NGC 2366  &  THINGS  &  07:28:53.4  &  +69:12:51  &  3.21  &  63  &  9.0  &  8.8  &  66.7  &  10  \\ 
Ho I  &  THINGS  &  09:40:32.3  &  +71:10:56  &  3.90  &  13  &  8.4  &  8.2  &  15.9  &  10  \\ 
NGC 3741  &  VLA-ANGST  &  11:36:06.4  &  +45:17:07  &  3.24  &  64  &  8.1  &  7.9  &  4.5  &  10  \\ 
Sextans A  &  VLA-ANGST  &  10:11:00.8  &  -04:41:34  &  1.38  &  36  &  8.0  &  7.8  &  4.3  &  10  \\ 
DDO 53  &  THINGS  &  08:34:07.2  &  +66:10:54  &  3.61  &  27  &  8.0  &  7.8  &  3.4  &  10  \\ 
DDO 190  &  VLA-ANGST  &  14:24:43.5  &  +44:31:33  &  2.79  &  30  &  8.0  &  7.6  &  7.9  &  10  \\ 
DDO 125  &  VLA-ANGST  &  12:27:41.8  &  +43:29:38  &  2.58  &  63  &  7.9  &  7.5  &  6.7  &  10  \\ 
Sextans B  &  VLA-ANGST  &  10:00:00.1  &  +05:19:56  &  1.39  &  30  &  7.9  &  7.6  &  1.8  &  10  \\ 
DDO 99  &  VLA-ANGST  &  11:50:53.0  &  +38:52:50  &  2.59  &  60  &  7.9  &  7.7  &  5.2  &  10  \\ 
UGCA 292  &  VLA-ANGST  &  12:38:40.0  &  +32:46:00  &  3.62  &  16  &  7.8  &  7.6  &  3.3  &  10  \\ 
NGC 4163  &  VLA-ANGST  &  12:12:09.1  &  +36:10:09  &  2.86  &  45  &  7.7  &  7.0  &  2.4  &  10  \\ 
UGC 4483  &  VLA-ANGST  &  08:37:03.0  &  +69:46:31  &  3.41  &  42  &  7.7  &  7.5  &  3.2  &  10  \\ 
DDO 181  &  VLA-ANGST  &  13:39:53.8  &  +40:44:21  &  3.14  &  50  &  7.7  &  7.4  &  3.5  &  10  \\ 
UGC 8833  &  VLA-ANGST  &  13:54:48.7  &  +35:50:15  &  3.08  &  33  &  7.4  &  7.1  &  1.7  &  10  \\ 
DDO 187  &  VLA-ANGST  &  14:15:56.5  &  +23:03:19  &  2.21  &  55  &  7.3  &  7.1  &  1.3  &  10  \\ 
GR 8  &  VLA-ANGST  &  12:58:40.4  &  +14:13:03  &  2.08  &  33  &  7.1  &  6.8  &  1.3  &  10  \\ 
\enddata

\tablecomments{The sample. Galaxies are listed in order of decreasing \m{baryon,tot}. All references are as given in \citetalias{StilpGlobal}. (1) Galaxy name; (2) \hi{} survey; (3-4) Position in J2000 coordinates; (5) Distance in Mpc; (6) Inclination; (7) \m{baryon, tot} in log \msun{}; (8) \m{HI, tot} in log \msun{}; (9) ANGST SFR$_\star (< 100 \; \mathrm{Myr})$; (10) de Vaucouleurs T-type.}

\end{deluxetable}}
\fi
%\FloatBarrier

\subsection{SF Histories}
\label{angst::sec:sample--sfhs}

To determine the time-resolved SFHs, we use data from ANGST, which provides multi-color \emph{Hubble Space Telescope} photometry of resolved stars in 69 nearby galaxies.
The survey and data processing pipeline are described in more detail in \citet{Dalcanton2009}.
The calibrated photometric data from ANGST can be translated into color-magnitude diagrams (CMDs), which can then be modeled to estimate the time-resolved SFH of the constituent stars \citeeg{Dolphin2002}.
As described in \citet{Dolphin2002}, the SFHs are generated by modeling each CMD as the linear combination of simple stellar populations with a variety of ages, assuming a single power law IMF $dN / dm \propto m^{-2.3}$.
When calculating the SFHs, other effects such as dust reddening, completeness limits, and photometric errors are taken into account.
The uncertainties on the SFHs for each galaxy are estimated for each galaxy by Monte Carlo realizations of the SFH, which account for uncertainty in stochastic sampling of the IMF. % and in the stellar models.
Further details on the generation of the SFHs used this paper can be found in \citet{Weisz2011}.

\begin{figure}
\ifthesis
%\begin{fullpage}
\fi
\centering
\includegraphics[width=6in]{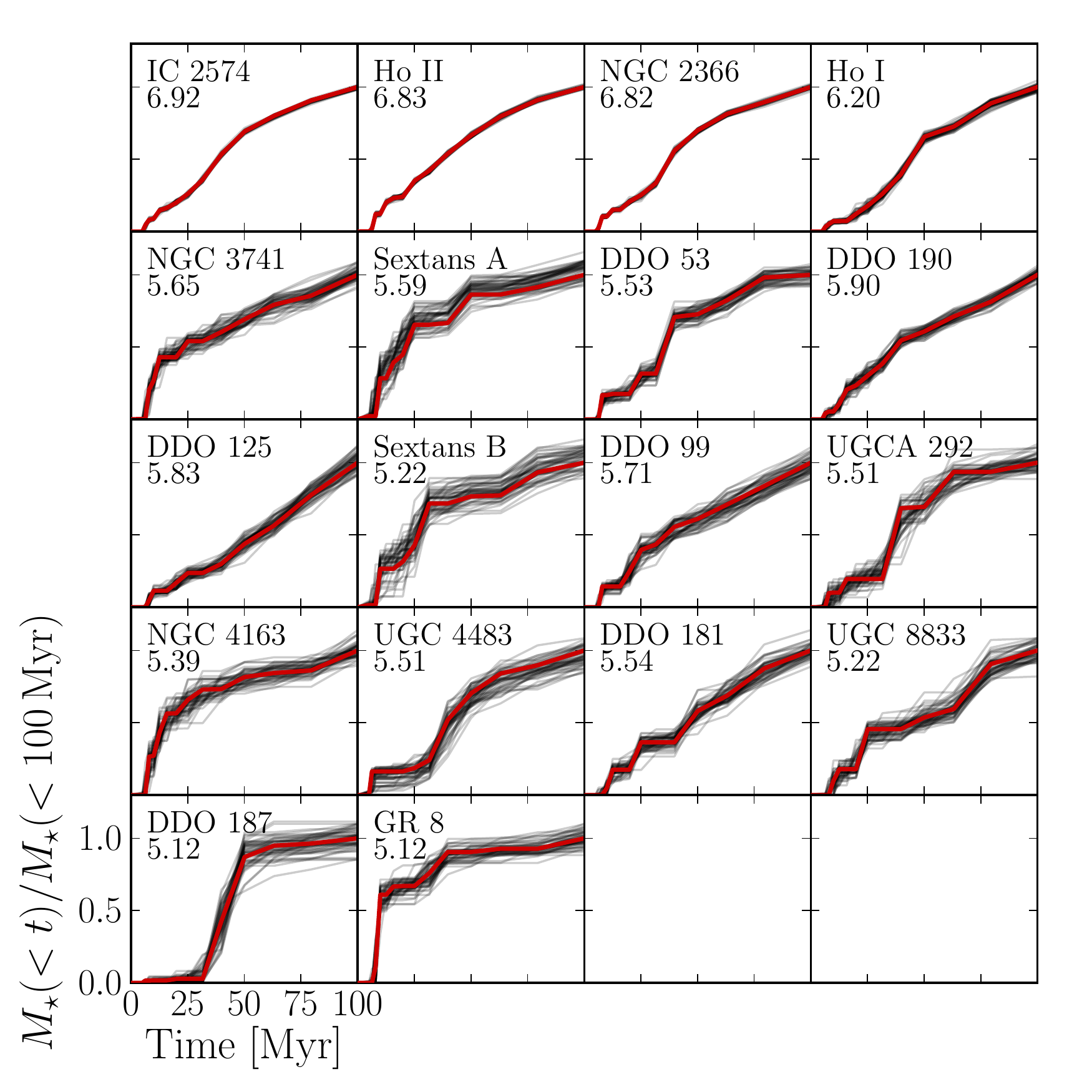}
\caption[Cumulative SFHs for the sample]{The cumulative number of stars formed, \mstarcumul{t}, as a function of time for the sample.
Each panel shows one galaxy.
The best fit SFH to the CMD shown in red.
The transparent black lines represent different Monte Carlo realizations of the SFH with appropriate uncertainties.
The $y$-axis is normalized to \mstarcumul{100~\mathrm{Myr}} for all galaxies, and the MC runs are scaled to \mstarcumul{100~\mathrm{Myr}} of the best-fit SFH.
The value of $\log \mstarcumul{100~\mathrm{Myr}}$ is shown below the galaxy name, in units of $\log \msun{}$.
\label{angst::fig:sfhs} }
\ifthesis
%\end{fullpage}
\fi
\end{figure}

The time-resolved cumulative SFHs for the sample are shown in Figure~\ref{angst::fig:sfhs}, in order of decreasing \m{baryon,tot}.
Each panel represents a single galaxy.
Within one panel, the thick red line shows the best-fit cumulative mass in stars formed between now and some time $t$ in the past (\mstarcumul{t}), normalized to the total mass in stars formed within the past 100~Myr (i.e., $\mstarcumul{100~\mathrm{Myr}}$); the value of $\log \mstarcumul{100~\mathrm{Myr}}$ for each galaxy is shown beneath the galaxy name.
The transparent black lines show the Monte Carlo realizations of the SFH, scaled to $\mstarcumul{100~\mathrm{Myr}}$ of the best-fit SFH.
Galaxies with smaller scatter around the best-fit line are therefore more certain.
Some galaxies (e.g., DDO~125 and DDO~190) show relatively constant SFHs, characterized by a straight diagonal line, while others have either more SF at recent times (e.g., Sextans~A, GR~8) or little recent SF (e.g., DDO~187, UGC~4483).
Figure \ref{angst::fig:sfhs} gives the impression that the variations away from a constant SFR increase in significance with decreasing \m{baryon,tot}.

The SFRs determined by ANGST are often higher than those measured by the FUV + 24\um{} method described in \citetalias{StilpGlobal}.
This discrepancy has been previously noted by \citet{Johnson2012}, who found that non-uniform SFHs can introduce a factor of $\sim 2$ scatter to the SFR as determined by FUV emission, as well as a systematic offset in the spectral synthesis models used to calibrate FUV SFR indicators.

\subsection{\hi{}}
\label{angst::sec:sample--hi}

We use \hi{} data cubes from THINGS and VLA-ANGST to estimate the \hi{} kinematics and kinetic energies of the sample.
We use the same data processing as described in \citetalias{StilpGlobal}.
To briefly summarize, we produce data cubes smoothed to 200~pc physical resolution using the AIPS task \textsc{convl}.
We then regenerate masks by blanking noise-only regions and recalculate moment maps using the convolved cubes and associated masks.
We note that the moment maps are generated from the flux-rescaled cubes, following \citet{Walter2008} and \citet{Ott2012}.
We use the 200 pc resolution data cubes and moment maps throughout the remainder of this paper.

\section{Analysis}
\label{angst::sec:analysis}

%In this section, we compare the \hi{} energies with the SFHs in the sample galaxies.
In this section, we describe the measurements we use to compare time-resolved SF to the properties of the neutral ISM.
We first explain how we quantify both the \hi{} energies in \S~\ref{angst::sec:analysis--measurements-hi} and the time-resolved SFH in \S~\ref{angst::sec:analysis--measurements-sfh}.

\subsection{\hi{} Energies}
\label{angst::sec:analysis--measurements-hi}

For all measurements of \hi{} energies, we include only those pixels above a signal-to-noise ratio $S/N > 5$, where $S/N$ is determined as the ratio between the maximum of the \gh{} polynomial fit to the line-of-sight spectrum of that pixel and the noise in a single channel, $\sigma_\mathrm{chan}$.
To generate a matched-aperture measurement, we also select only those pixels that are also in the aperture of the ANGST observations.
The included pixels for each galaxy are those in Figure~\ref{angst::fig:footprints} that are within both the black ANGST footprints and the red $S/N > 5$ contours.

\begin{figure}
\centering
\includegraphics[width=6in]{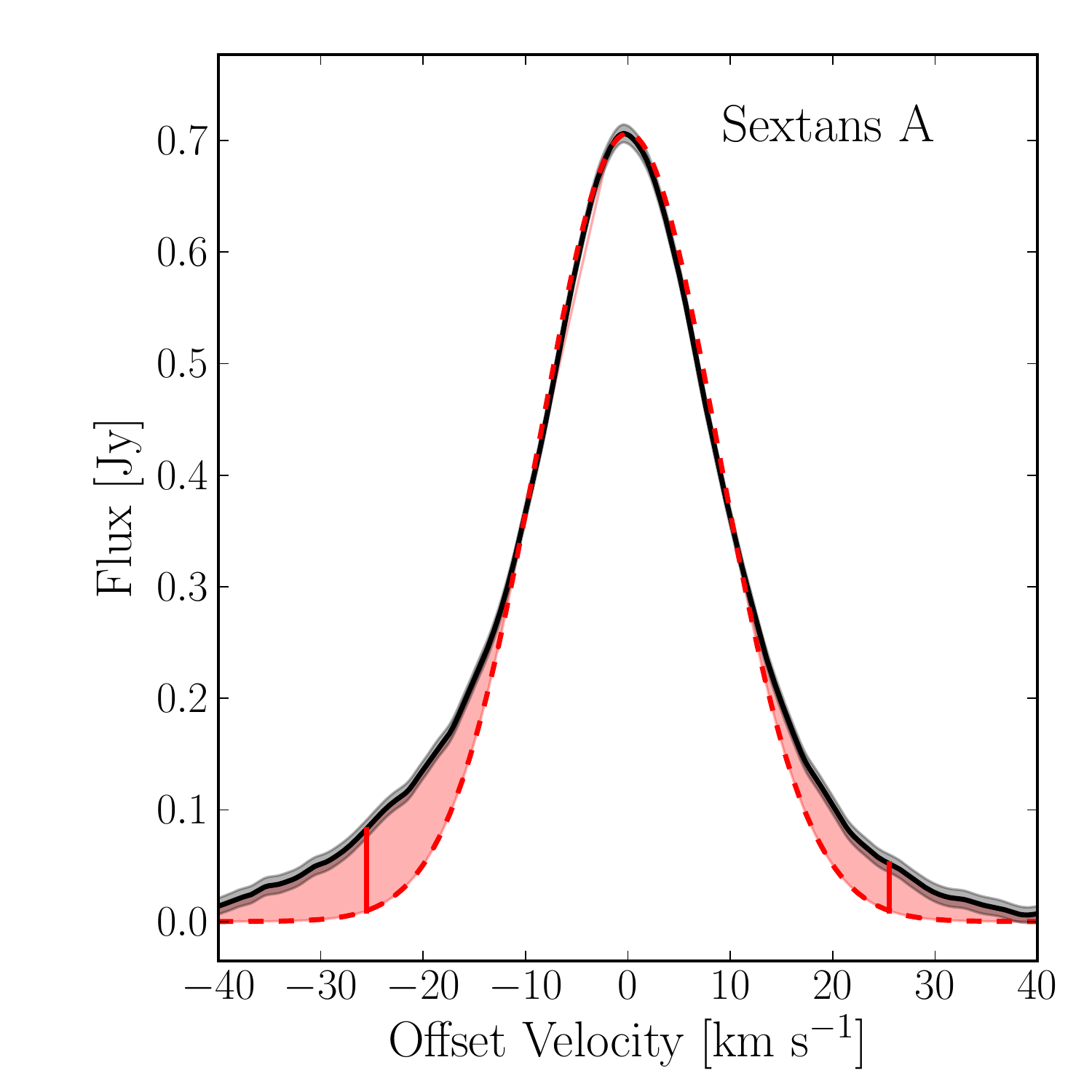}
\caption[The superprofile for Sextans A in the ANGST FOV]{The superprofile for Sextans~A in the ANGST FOV.
The thick black line is the superprofile itself, with noise shown as the shaded grey region surrounding the superprofile.
The thick dashed red line is the HWHM-scaled Gaussian model.
The pink shaded region between the model and the superprofile represents \fw{}, and the solid red vertical lines show $\pm \swing{}$.
\label{angst::fig:sexa-superprofile}}
\end{figure}

Our analysis is based on \hi{} superprofiles using the methodology presented in \citetalias{StilpGlobal}, where we have removed the relative line-of-sight velocity (largely due to rotation) and co-added the resulting line-of-sight spectra to measure the integrated turbulent velocity structure with high signal-to-noise.
To derive the superprofiles, we determine the velocity of the peak (\vp{}) of the line-of-sight spectrum along each pixel, for all pixels also in the ANGST aperture.
We then co-add these line-of-sight spectra after subtracting each pixel's \vp{} to produce a flux-weighted average \hi{} line profile.
An example superprofile for Sextans A is shown in Figure~\ref{angst::fig:sexa-superprofile}.
The superprofile itself is shown as the thick black line, and the uncertainty is shown as the grey shaded region around the superprofile.

We model the central peak of the superprofile with a Gaussian profile scaled to its half-width half-maximum (HWHM) and amplitude, and adopt the parameterization and physical interpretation presented in \citetalias{StilpGlobal}.
Specifically, the Gaussian width that corresponds to the HWHM model yields \scentral{} and represents the average turbulent velocity of the \hi{}.
Gas with velocities above that expected from a Gaussian core is referred to as the ``wings.''
This anomalous gas is potentially due to expanding or asymmetric \hi{} structures but may also be an inherent property of \hi{} line profiles (e.g., \citetalias{StilpGlobal}).

We calculate \fw{}, the fraction of gas moving faster than expected based on the HWHM Gaussian model, as
 \begin{equation}
\fw{} = \frac{ \sumlim{|v| > \mathrm{HWHM}} \left[ S(v) - G(v) \right]} { \sumlim{|v| > 0} S (v) }.
\label{angst::eqn:fw}
\end{equation}
In this equation, $v$ is the velocity offset relative to the peak. 
$S(v)$ and $G(v)$ are the measured superprofile and HWHM Gaussian model, respectively.
We represent the flux contributing to \fw{} as the transparent shaded red region in Figure~\ref{angst::fig:sexa-superprofile}.

We also quantify the \emph{rms} velocity of the excess \hi{} in the wings:
\begin{equation}
\swingsq{} = \frac{ \sumlim{ | v | > \mathrm{HWHM}}  \left[ S (v) - G(v) \right] v^2 }  { \sumlim{ | v | > \mathrm{HWHM}} \left[ S (v) - G(v) \right] } .
\label{angst::eqn:swing}
\end{equation}
This parameter is proportional to the energy per unit mass in the wings of the superprofile.
In Figure~\ref{angst::fig:sexa-superprofile}, we show \swing{} as a solid vertical red line at $\pm \swing{}$.

For each galaxy, we also characterize the average line width of the \hi{} width using the average of the second moment value of the velocity, $\sigma$.
The second moment is often used in the literature and provides an estimate of \hi{} kinematics that is independent of the superprofile parameterization chosen above.
This quantity makes no assumption about the underlying structure of \hi{} emission \citeeg{Tamburro2009}.
We use the second moment values of all included pixels to calculate a flux-weighted average second moment:
\begin{equation}
\smomtwo{} = \frac{\sumlim{i} \sigma_{i} N_{\mathrm{HI},i}}{\sumlim{i} N_{\mathrm{HI},i}},
\end{equation}
where $N_{\mathrm{HI},i}$ is the column density of each pixel included in the average and $\sigma_i$ is the second velocity moment of each line-of-sight spectrum.
The flux weighting in this equation accounts for the fact that regions of high \hi{} column density require more energy input compared to regions of low \hi{} column density to produce the same line width.

We list the three \hi{} velocities for the sample in Table~\ref{tab:angst--parameters}, in order of decreasing \m{baryon,tot}.
The columns are
(1)~galaxy name;
(2)~velocity resolution, $\Delta v$,
(3)~the number of independent resolution elements contributing to the superprofile, $N_\mathrm{beams}$;
(4)~the average \hi{} surface density associated with the superprofile, $\m{SP} / A_\mathrm{HI}$;
(5)~\smomtwo{};
(6)~\scentral{};
(7)~\swing{};
(8)~\fw{}.

\ifthesis
\afterpage{\input{\tablepath/table-parameters}}
\else
\afterpage{\begin{deluxetable}{lccccccc}
\tablewidth{0pt}
\tabletypesize{\scriptsize}
\tablecaption{ANGST superprofile parameters. \label{tab:angst--parameters}}

\tablehead{

  \colhead{1} &
  \colhead{2} &
  \colhead{3} &
  \colhead{4} &
  \colhead{5} &
  \colhead{6} &
  \colhead{7} &
  \colhead{8} \\

  \colhead{Galaxy} &
  \colhead{$\Delta v$} & 
  \colhead{$N_\mathrm{beams}$}  & 
  \colhead{$\langle \m{HI} / A_\mathrm{HI} \rangle$} & 
  \colhead{\smomtwo{}} &
  \colhead{\scentral{}} & 
  \colhead{\swing{}} & 
  \colhead{\fw{}} \\
  
  \colhead{} & 
  \colhead{(\kms{})} & 
  \colhead{} & 
  \colhead{(\msun{} \per[2]{pc})} & 
  \colhead{(\kms{})} & 
  \colhead{(\kms{})} & 
  \colhead{(\kms{})} & 
  \colhead{} \\

}

\startdata
IC 2574  & 2.6  &  810   &  5.9  &  $ 9.9 \pm  1.7$  &  $ 7.6 \pm  0.5$  &  $21.9 \pm  0.7$  &  $0.12 \pm 0.01$  \\ 
Ho II  & 2.6  &  281   &  7.3  &  $10.4 \pm  1.7$  &  $ 7.8 \pm  0.6$  &  $23.7 \pm  1.5$  &  $0.14 \pm 0.01$  \\ 
NGC 2366  & 2.6  &  433   &  7.6  &  $13.8 \pm  2.0$  &  $10.3 \pm  0.5$  &  $31.2 \pm  0.9$  &  $0.13 \pm 0.01$  \\ 
Ho I  & 2.6  &  152   &  9.6  &  $ 8.8 \pm  1.2$  &  $ 6.9 \pm  0.6$  &  $21.3 \pm  2.1$  &  $0.13 \pm 0.02$  \\ 
NGC 3741  & 1.3  &   85   &  3.7  &  $ 9.3 \pm  1.6$  &  $ 7.4 \pm  0.6$  &  $20.0 \pm  2.4$  &  $0.12 \pm 0.02$  \\ 
Sextans A  & 1.3  &   17   &  7.7  &  $10.9 \pm  2.3$  &  $ 8.7 \pm  0.6$  &  $25.6 \pm  1.9$  &  $0.14 \pm 0.02$  \\ 
DDO 53  & 2.6  &   92   &  8.4  &  $10.0 \pm  1.3$  &  $ 8.2 \pm  0.6$  &  $23.8 \pm  1.6$  &  $0.11 \pm 0.02$  \\ 
DDO 190  & 2.6  &   64   &  8.7  &  $10.6 \pm  1.2$  &  $ 9.1 \pm  0.7$  &  $25.3 \pm  2.6$  &  $0.07 \pm 0.01$  \\ 
DDO 125  & 0.6  &   39   &  2.9  &  $ 7.3 \pm  0.7$  &  $ 6.1 \pm  0.6$  &  $17.0 \pm  2.7$  &  $0.15 \pm 0.03$  \\ 
Sextans B  & 1.3  &   18   &  5.9  &  $ 9.0 \pm  1.0$  &  $ 7.8 \pm  0.6$  &  $21.2 \pm  1.9$  &  $0.10 \pm 0.01$  \\ 
DDO 99  & 1.3  &   37   &  3.5  &  $ 8.5 \pm  1.4$  &  $ 7.1 \pm  0.6$  &  $19.1 \pm  2.0$  &  $0.11 \pm 0.02$  \\ 
UGCA 292  & 0.6  &   45   & 13.3  &  $ 8.6 \pm  0.7$  &  $ 7.8 \pm  0.6$  &  $20.3 \pm  3.5$  &  $0.06 \pm 0.02$  \\ 
NGC 4163  & 0.6  &   14   &  5.5  &  $ 8.4 \pm  0.6$  &  $ 7.7 \pm  0.7$  &  $22.1 \pm  3.0$  &  $0.09 \pm 0.04$  \\ 
UGC 4483  & 2.6  &   43   &  7.0  &  $10.1 \pm  1.1$  &  $ 8.3 \pm  0.6$  &  $23.8 \pm  2.6$  &  $0.10 \pm 0.02$  \\ 
DDO 181  & 1.3  &   46   &  4.0  &  $ 8.1 \pm  0.7$  &  $ 6.6 \pm  0.6$  &  $19.9 \pm  2.7$  &  $0.13 \pm 0.03$  \\ 
UGC 8833  & 2.6  &   30   &  6.1  &  $ 9.8 \pm  1.1$  &  $ 8.0 \pm  0.7$  &  $23.2 \pm  2.8$  &  $0.11 \pm 0.02$  \\ 
DDO 187  & 1.3  &   20   &  5.3  &  $11.6 \pm  1.5$  &  $10.6 \pm  0.7$  &  $25.2 \pm  3.2$  &  $0.09 \pm 0.02$  \\ 
GR 8  & 0.6  &   15   &  4.7  &  $ 8.0 \pm  0.5$  &  $ 7.5 \pm  0.6$  &  $20.0 \pm  2.4$  &  $0.12 \pm 0.03$  \\ 
\enddata

\tablecomments{Sample parameters. Galaxies are listed in order of decreasing \m{baryon,tot}. (1) Galaxy name. (2) Velocity resolution. (3) $N_\mathrm{beams}$ of the superprofiles. (4) Average \hi{} surface density of the superprofiles. (5) \smomtwo{}. (6) \scentral{}. (7) \swing{}. (8) \fw{}. }

\end{deluxetable}
}
\fi
%\clearpage

We use three \hi{} energy estimates for our analysis:
one based on the central peak of the superprofiles (corresponding to \scentral{}),
one based on the wings of the superprofiles (corresponding to \swing{}),
and one based on the average second moment value for each galaxy (corresponding to \smomtwo{}).
However, these velocities alone do not provide an ideal comparison with energy input from SF, because regions with the same \hi{} energy but different \hi{} masses will also have different line widths.
We therefore use the average energy in the superprofile peak, the superprofile wings, and the entire line profiles when comparing to SF.

We estimate the energy surface density in the central peak of the superprofile as:
\begin{equation}
\Sigma_\mathrm{E,central} = \frac{3}{2} \frac{\m{SP}}{A_\mathrm{HI}} (1 - \fw{}) (1 - \fcold{}) \sigma_\mathrm{central}^2,
\label{angst::eqn:e-central}
\end{equation}
where $\m{SP} (1 - \fw{})$ is the total \hi{} mass contained in the central peak, and $A_\mathrm{HI}$ is the area covered by the \hi{}.
The $(1 - \fcold{})$ correction accounts for the mass of dynamically cold \hi{} ($\sigma{} < 6$ \kms{}), which has kinematics that are not well-described by \scentral{}.
While the fraction of cold \hi{} is very uncertain in dwarfs, we choose $\fcold = 0.15$, a value in line with previous estimates of \fcold{} in  dwarf galaxies \cite[between $1 - 20$\%;][]{Young2003, Bolatto2011, Warren2012}.
The contribution of cold \hi{} gas to the superprofile may account for the fact that the central peaks are narrower than the HWHM Gaussian model.
The $3/2$ factor accounts for motion in all three directions, assuming an isotropic velocity dispersion.

Second, we estimate the energy surface density in the wings of the superprofile as:
\begin{equation}
\Sigma_\mathrm{E,wings} = \frac{3}{2} \frac{\m{SP}}{A_\mathrm{HI}} 
  \fw{} \sigma_\mathrm{wings}^2.
\label{angst::eqn:e-wings}
\end{equation}
Here, $\left( \m{SP} / A_\mathrm{HI} \right) \times \fw{}$ 
represents the total \hi{} surface density associated with the wings of the superprofile.

Finally, we estimate the average energy surface density of the entire line-of-sight profiles as:
\begin{equation}
\Sigma_\mathrm{E,m2} = \frac{3}{2} \frac{\m{SP}}{A_\mathrm{HI}} %\Sigma_\mathrm{HI}
\sigma_\mathrm{m2}^2,
\label{angst::eqn:e-m2}
\end{equation}
where $\m{SP} / A_\mathrm{HI}$ is the average \hi{} surface density of the superprofile.
This estimate accounts for the energy in the full line profile and is independent of parameterization, and essentially combines the energies in Equations~\ref{angst::eqn:e-central} and \ref{angst::eqn:e-wings}.

\subsection{SFH Measurements}
\label{angst::sec:analysis--measurements-sfh}

Next, we measure the time-resolved SFRs.
We use the ANGST SFHs to calculate the average SFR between times $t_i$ and $t_f$:
\begin{equation}
\avesfr{t_i \rightarrow t_f} = \frac{\mstarcumul{t_f} - \mstarcumul{t_i}}{t_f - t_i},
\end{equation}
where $\mstarcumul{t}$ is the total stellar mass formed between now and a time $t$ in the past, within the ANGST aperture.
Note that $t_i$ and $t_f$ are the initial and final times in an integration going back in time, where the present is defined as $t = 0$ and \emph{not} the initial and final times of a SF event.
The SFHs derived from ANGST are uniformly spaced in logarithmic time intervals, with $\Delta \log_{10} (t / \mathrm{Myr}) = 0.1$.
For our analysis, however, we work in linearly-spaced 10~Myr bins, a value that is well-matched to the theoretically dissipation timescale of turbulence in the ISM \citep{MacLow1999}.
To find the value of \mstarcumul{t}, we assume that the star formation rate has been constant over each time bin, and linearly interpolate between the bin edges to find \mstarcumul{t}.
In some cases, both $t_i$ and $t_f$ fall in the same bin, which has the effect of smoothing out SFR variations at times further in the past.
However, our results do not change if we instead use the intrinsic logarithmic time bins instead of the 10~Myr linear bins.

%This interpolation has a larger effect at times further in the past because the input time bins are uniformly spaced in logarithmic time with $\Delta \log_{10} (t / \mathrm{Myr}) = 0.1$.
%We adopt a time step of 10~Myr, a value that is well-matched to the theoretical dissipation timescale of turbulence in the ISM \citep{MacLow1999}.
%However, our results do not change if we instead use the intrinsic logarithmic time bins instead of the 10~Myr linear bins.

From \avesfr{t_i \rightarrow t_f}, we can calculate the SFR surface density over that time range, \avesfrsd{t_i \rightarrow t_f}.
For this measurement, we divide \avesfr{t_i \rightarrow t_f} by $A_\mathrm{HI}$, corrected for galaxy inclination.
Some of the dwarf galaxies are smaller than the ANGST aperture, but in these cases the SF is primarily contained in the same regions as the \hi{}.

\section{Comparing \hi{} Energy to Time-Resolved SF}
\label{angst::sec:comparisons}

In this section, we compare the \hi{} energies to the time-resolved \sfrsd{} values.
If the \hi{} turbulent energy is supplied by SF, then we would expect to see a correlation between \sfrsd{} and at least one measure of \hi{} kinetic energy.
However, the correlation may depend on the time interval being considered.
We can only  measure the \hi{} kinematics at the present time, but the relevant SF energy driving the turbulence is not necessarily from the most recent SF.
Instead, the timescale for energy input may reflect the variations in the SFR with time, the numbers and masses of evolving stars at each time, and the timescale for SNe and stellar winds to couple to the neutral ISM.
A strong correlation between \hi{} energy and SF at a specific timescale would support the idea that SF and \hi{} kinematics are coupled on that timescale.

In \S~\ref{angst::sec:comparisons--uncertainties} we discuss our method for measuring correlations and deriving the associated uncertainties.
In \S~\ref{angst::sec:comparisons--correlations-tf}, we examine correlations in the mean \sfrsd{} measured between now (i.e., $t_i$ = 0) and several $t_f$ values.
In \S~\ref{angst::sec:comparisons--correlations-ti-tf}, we search for correlations between all possible pairs of $t_i$ and $t_f$ between now and 100~Myr in the past.

\subsection{The Spearman Correlation Coefficient \rs{} and Associated Uncertainties}
\label{angst::sec:comparisons--uncertainties}

We measure the degree of correlation between \hi{} kinetic energy and SFR on a given timescale using the Spearman rank correlation coefficient, \rs{}. This statistic tests only for a monotonic relationship between the two input data sets.
The statistic yields $0 < \rs{} \leq 1$ for a positive correlation, $\rs{} = 0$ for completely uncorrelated data, and $-1 \leq \rs{} < 0$ for an anticorrelation.
The probability of finding an \rs{} value equal to or more extreme than measured from a random data set is given by \ps{}.

To interpret the significance of \rs{}, we must have a reliable estimate of the associated uncertainties.
There are two sources of uncertainty on the measured \rs{} values.
First, uncertainties in the data themselves can propagate to uncertainties in \rs{}.
Second, the small number of galaxies in this study may not adequately sample the parameter space, potentially skewing \rs{} values.
In this section, we assess the uncertainty due to each of these factors.

We first estimate the uncertainties on \rs{} due to uncertainties in the data themselves.
For the \hi{} \smomtwo{} value, we adopt as the uncertainty the flux-weighted standard deviation of the second moment values for the included pixels.
This estimate provides a measurement of the spread of observed second moment values contributing to each superprofile.
For the \hi{} kinematic measurements that contribute to $\Sigma_\mathrm{E,central}$ and $\Sigma_\mathrm{E,wings}$, we approximate the uncertainties on the measured superprofile parameters based on the noise on each point, given by:
\begin{equation}
\sigma_\mathrm{SP} = \sigma_\mathrm{chan} 
\times \sqrt{ N_\mathrm{pix} / N_\mathrm{pix/beam}}
\times \frac{F_\mathrm{rescaled}}{F_\mathrm{standard}},
\label{angst::eqn:noise-jvm}
\end{equation}
where $\sigma_\mathrm{chan}$ is the \emph{rms} noise in a single channel; $N_\mathrm{pix}$ is the number of channels contributing to a superprofile point, and $N_\mathrm{pix/beam}$ is the number of pixels per resolution element; and $F_\mathrm{rescaled} / F_\mathrm{standard}$ the flux ratio between the total measured flux in the superprofile generated from the flux-rescaled cube to that from the standard cube.
This factor approximates the rescaling that accounts for the difference in beam area between the residuals and the clean components from deconvolution.
These uncertainties are explained in more detail in \citetalias{StilpGlobal}.
We assume that the observed superprofile is correct, and add Gaussian noise to each point based on Equation~\ref{angst::eqn:noise-jvm}.
We then calculate ``noisy'' parameters for this measurement.
After repeating this process 1,000 times, we have determined the allowed distribution of superprofile parameters due to noise.
We fit a Gaussian to each parameter's distribution of noisy values and adopt its $1 \sigma$ width as the uncertainty on that parameter.
We also include systematic uncertainties on the parameters based on finite velocity resolution, as described in detail in \citetalias{StilpGlobal}.
The uncertainties on the \hi{} parameters are given in Table~\ref{tab:angst--parameters}.

Uncertainties in the SFHs are more complicated, as neighboring time bins are correlated.
A burst of SF in one bin could actually have occurred in an adjacent time bin.
These uncertainties are accounted for in the Monte Carlo realizations shown in Figure~\ref{angst::fig:sfhs}.

We determine the uncertainties on \rs{} due to parameter uncertainties with a series of 1,000 realizations of our sample.
For each realization, we add an offset to each \hi{} kinematic parameter drawn from a Gaussian distribution with that parameter's uncertainty as the Gaussian standard deviation.
For each galaxy, we calculate \avesfrsd{t_i \rightarrow t_f} from a randomly-chosen Monte Carlo instance of the SFH (black lines in Figure~\ref{angst::fig:sfhs}; see \S~\ref{angst::sec:sample--sfhs}), instead of the best-fit SFH.
We then calculate the \rs{} value using the ``noisy'' \hi{} kinematic parameters and \avesfrsd{t_i \rightarrow t_f} from the random Monte Carlo SFH.
We then adopt the inner 68\% of all allowed \rs{} values as the uncertainty in \rs{} due to parameter uncertainties.

Second, the small number of points can affect \rs{} values if they are not adequately sampling the underlying distribution.
We account for this uncertainty using bootstrapping.
We randomly draw a new sample of the same size as our original sample, allowing repeats, and calculate \rs{} for the best-fit SFHs and parameters of the resample.
Repeating this procedure gives us a range of \rs{} values that are statistically allowed by the sample size.
As above, we adopt the inner 68\% of this range as the uncertainty in the \rs{} values.

The bootstrapped uncertainties in \rs{} due to the small sample size are typically larger than those due to uncertainties in the data.
For the rest of the paper, we assess the significance of measured \rs{} values using both uncertainty estimates individually.

\subsection{Correlations between \hi{} Energy and Mean SFR}
\label{angst::sec:comparisons--correlations-tf}

We start by comparing \hi{} energetics to SF surface density between now and some time $t_f$ in the past, \avesfrsd{0 \rightarrow t_f}, for $t$ between $10 - 100$ Myr in 10 Myr steps.
For each $t_f$ step, we calculate \rs{} between \avesfrsd{0 \rightarrow t} and each \hi{} energy parameter.
We then compare these correlation coefficients as a function of $t_f$ to identify the timescales for which the current energetics of the \hi{} gas are most coupled to SF.
We also calculate the uncertainties on \rs{} for each $t_f$ step based on the procedures described in \S~\ref{angst::sec:comparisons--uncertainties}.

\begin{figure}
\centering
\includegraphics[width=6in]{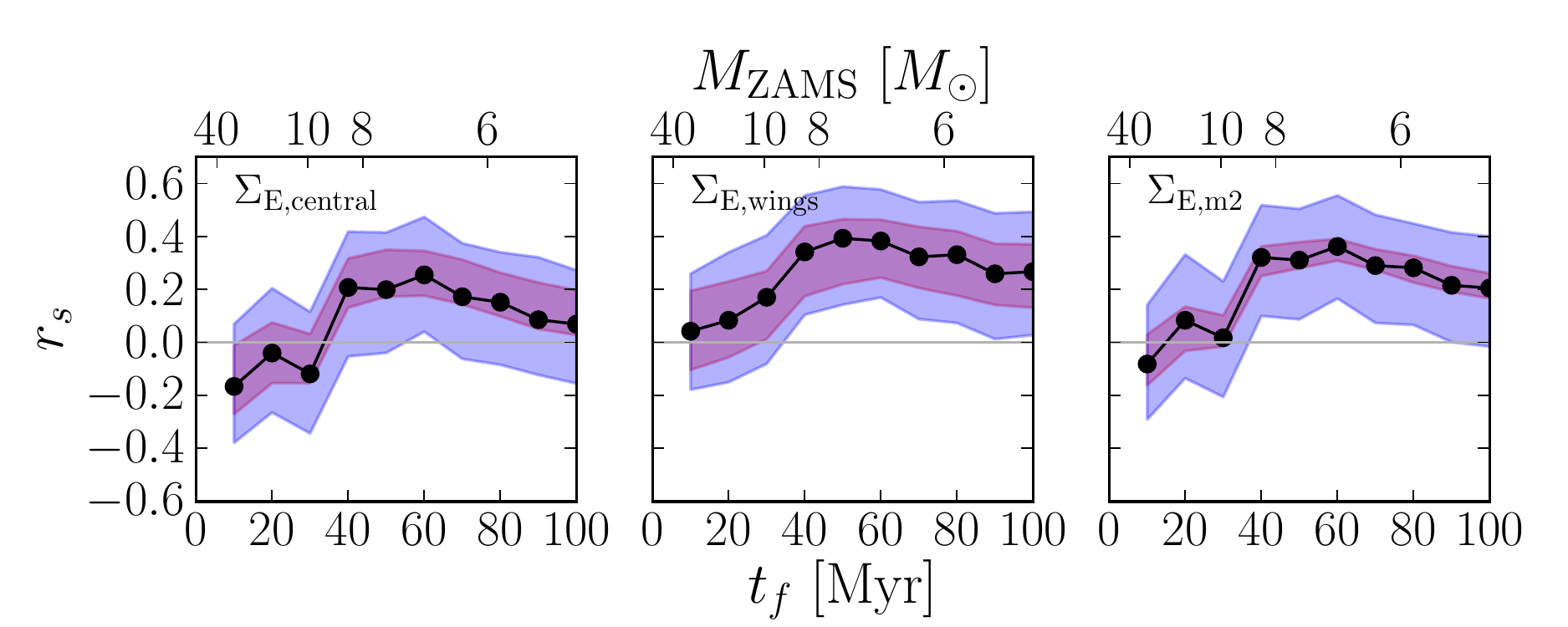}
\caption[Correlations between HI energy surface densities and mean \sfrsd{}]{Correlation coefficients (\rs{}) between \hi{} energy and \avesfrsd{0 \rightarrow t_f} for the entire sample.
We consider \hi{} energy in the central peak of the superprofiles (left panel), in the wings (right panel), and in the full line profiles (right panel).
The black points show the observed correlations data between the superprofiles and the best-fit SFHs averaged on timescales between $0 - t_f$.
The shaded red and blue regions indicate the allowed range in \rs{} from data uncertainties and bootstrapping.
We do not see positive correlations between \sfrsd{} and \hi{} energy until we include SF $\sim 40$~Myr in the past.
\label{angst::fig:corr-tf}
}
\end{figure}

In Figure~\ref{angst::fig:corr-tf}, we show the correlation coefficients between the three \hi{} energy parameters ($\Sigma_\mathrm{E,central}$, $\Sigma_\mathrm{E,wings}$, and $\Sigma_\mathrm{E,m2}$) and the integrated \avesfrsd{0 \rightarrow t_f} from the present to a lookback time $t_f$.
The black points represent correlations with the best-fit SFH, and the red and blue shaded regions indicate the allowed ranges in \rs{} due to the data uncertainties and bootstrapping, respectively.

We find that the \hi{} energies do not show significant correlations with the recent SFR ($t \leq 30$ Myr).
Even at $t_f \sim 40$~Myr, the correlation coefficients for the sample size are low ($r_s \lesssim 0.3$), implying that there is a $\sim 20$\% chance of drawing this sample from a random sample (i.e., $\ps{} \sim 0.2$).

\subsection{Correlations between \hi{} Energy and SF on Arbitrary Timescales}
\label{angst::sec:comparisons--correlations-ti-tf}

The method used in \S~\ref{angst::sec:comparisons--correlations-tf} assumes that all SF between $t_f$ and the present time affects the ISM.
However, the energy input from a SF burst is not necessarily constant or instantaneous.
Instead, the majority of energy is released some time after the SF burst actually occurs \citeeg{Leitherer1999}, when massive stars end their lives in SNe.
Furthermore, the neutral ISM may not show effects from SF until some time after the burst, due not only to this lag but also to a potential delay in converting the localized mechanical SF energy to global turbulent energy in the neutral ISM.
To address this issue, we compare the \hi{} energies with \avesfrsd{t_i \rightarrow t_f} for a range of $t_i$ and $t_f$ values, and generate an \rs{} value and associated uncertainties for each combination.

\begin{figure}
\centering
\begin{minipage}[b]{0.45\linewidth}
\centering
\includegraphics[width=2.8in]{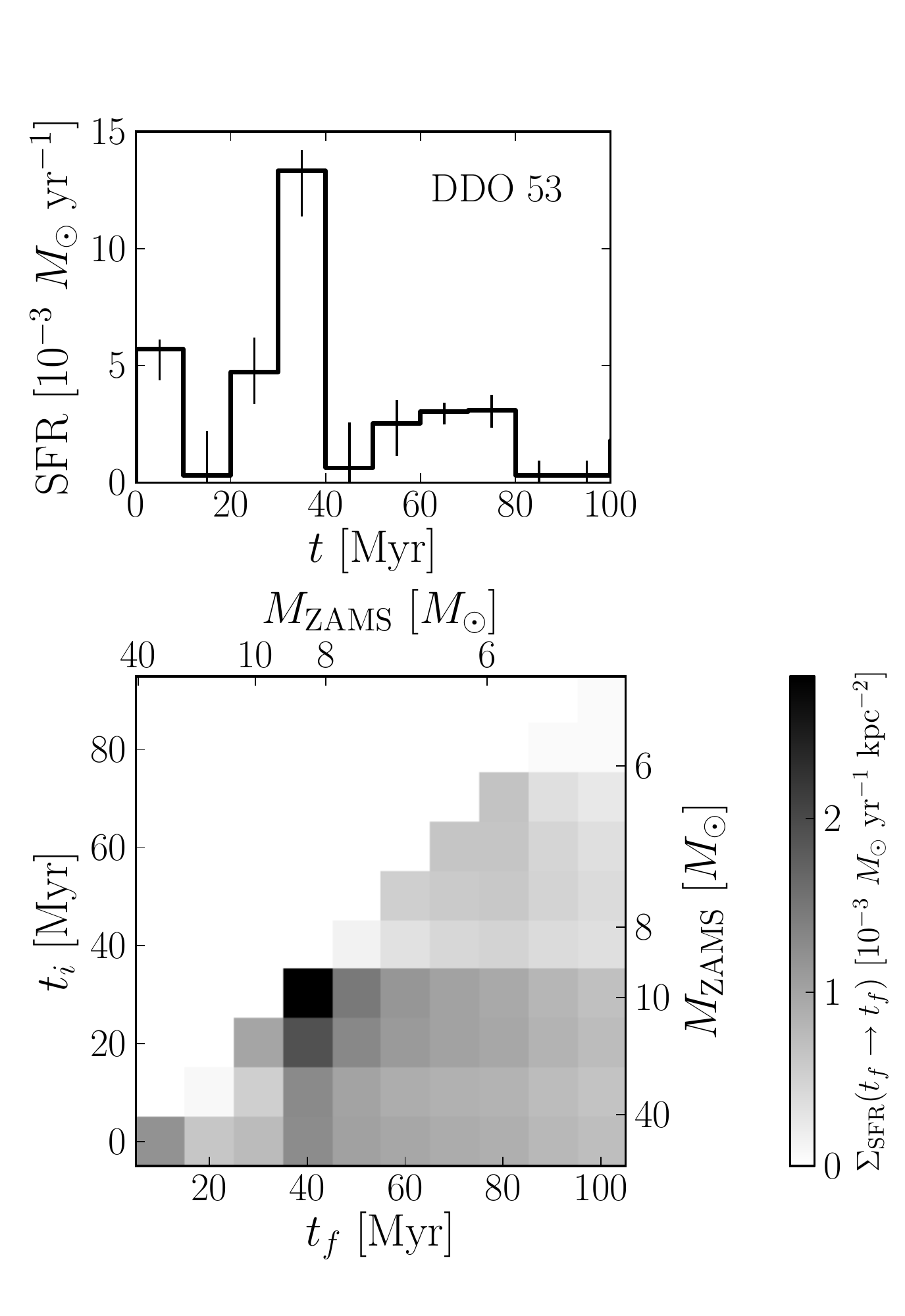}
\end{minipage}
\begin{minipage}[b]{0.45\linewidth}
\centering
\includegraphics[width=2.8in]{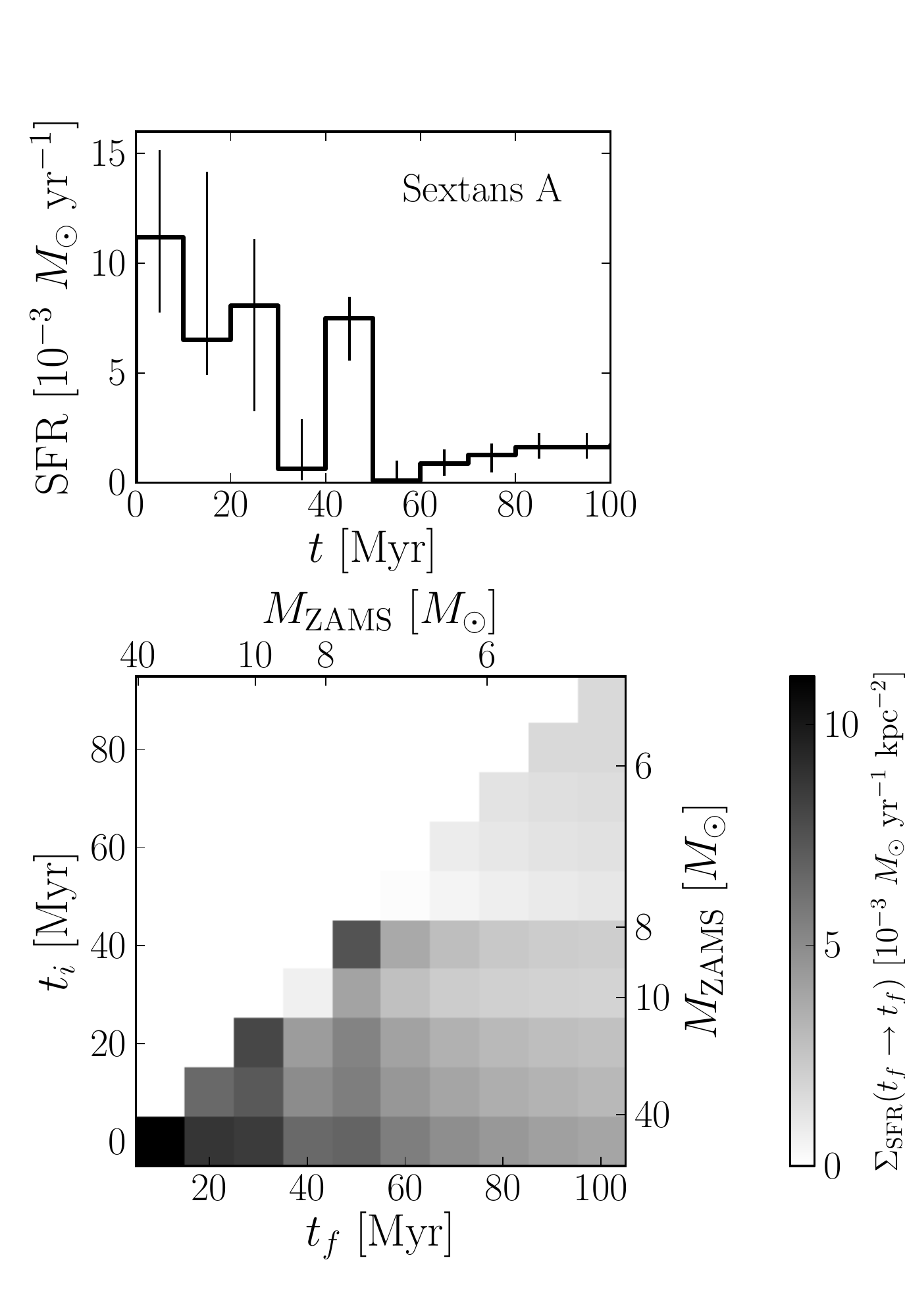}
\end{minipage}
\caption[Mean \sfrsd{} averaged over various timescales]{Mean \avesfrsd{t_i \rightarrow t_f} averaged over various $t_i$ and $t_f$ for two sample galaxies (DDO 53, left panels; Sextans A, right panels).
The upper panels show the total SFR measured in the ANGST aperture, averaged in 10~Myr time bins, with error bars derived from the MC realizations of the SFH.
The lower panels show the effects of averaging the measured \sfrsd{} from ANGST in different time ranges between $t_i$ and $t_f$.
The $y$-axis shows $t_i$, and the $x$-axis shows $t_f$.
The color of each box at a given $t_i$ and $t_f$ value represents \avesfrsd{t_i \rightarrow t_f}.
The boxes along the diagonal line are an alternative representation of the SFRs shown in the upper panels, after normalization by area (i.e., \sfrsd{} instead of SFR).
Large, short bursts of SF are smeared out as the $t_i \rightarrow t_f$ becomes larger.
\label{angst::fig:sfh-ti-tf}
}
\end{figure}

In Figure~\ref{angst::fig:sfh-ti-tf}, we first show the variation in \avesfrsd{t_i \rightarrow t_f} for different values of $t_i$ and $t_f$ for two sample galaxies.
The upper panels show the total SFR in the ANGST aperture, binned in 10~Myr intervals. 
The lower panels show \avesfrsd{t_i \rightarrow t_f}, where the $y$-axis shows the initial time $t_i$, the $x$-axis shows the final time $t_f$, and the greyscale in each panel indicates the SF surface density of the given time interval associated with that pixel.
Along the top and right axes, we have also shown the zero-age main sequence mass of the star whose lifetime corresponds to $t_i$ and $t_f$  ($M_\mathrm{ZAMS}$), using models from \citet{Marigo2008} and \citet{Girardi2010}.
Thus, above the one-to-one line is forbidden, and the boxes along the one-to-one line represents the shortest time interval (10~Myr averaging).
The color of each box represents \avesfrsd{t_i \rightarrow t_f} for the $t_i$ and $t_f$ corresponding to that box's position, for $t_i$ and $t_f$ between 0 and 100 Myr in 10 Myr steps.
It is clear that short bursts of SF can be smoothed out in larger $t_i \rightarrow t_f$ ranges, and in some cases, the average SFR can change by a factor of 10 based on the timescale considered.

We now consider the correlations in the entire sample between \hi{} kinetic energies and \sfrsd{}, averaged over the $t_i \rightarrow t_f$ ranges shown in Figure~\ref{angst::fig:sfh-ti-tf}.
For each $t_i \rightarrow t_f$ range, we calculate \avesfrsd{t_i \rightarrow t_f} for each galaxy, and calculate the \rs{} value between \avesfrsd{t_i \rightarrow t_f} and the \hi{} energy surface densities for the entire sample.

\begin{figure}
\centering
\includegraphics[width=4.5in]{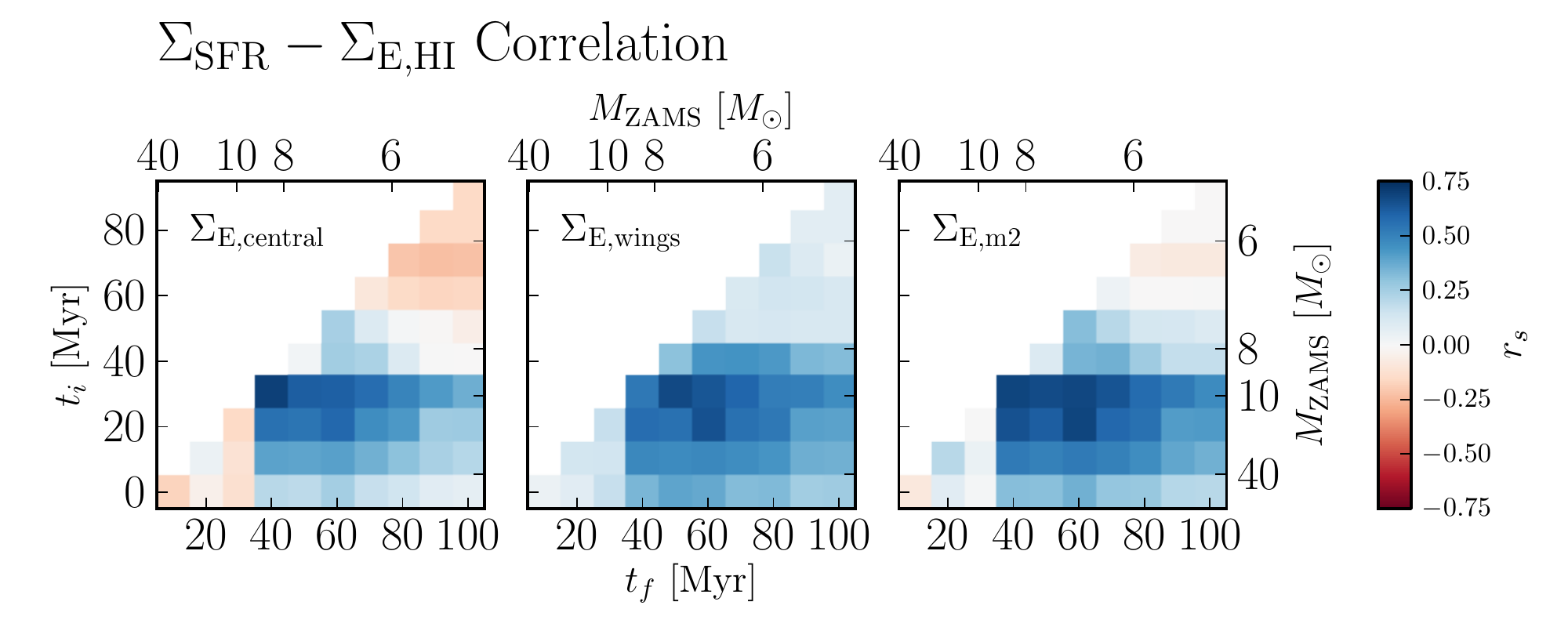}
\includegraphics[width=4.5in]{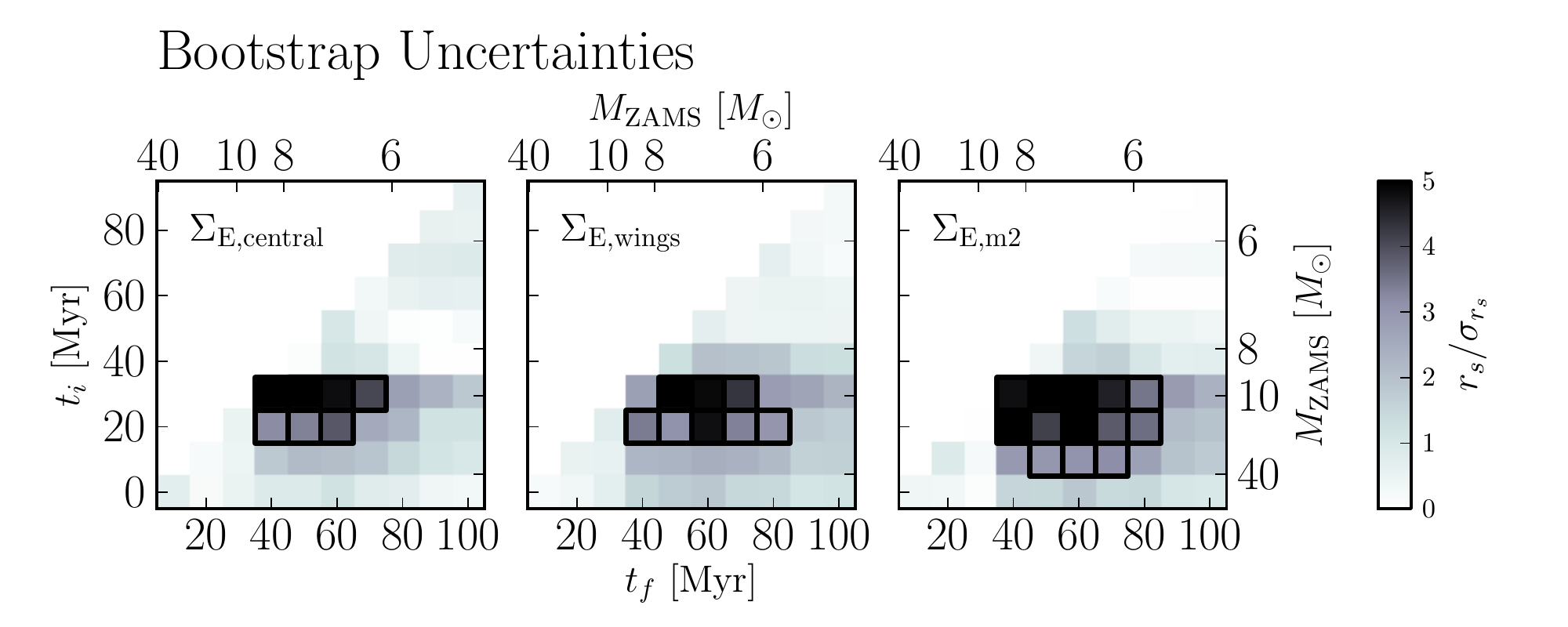}
\includegraphics[width=4.5in]{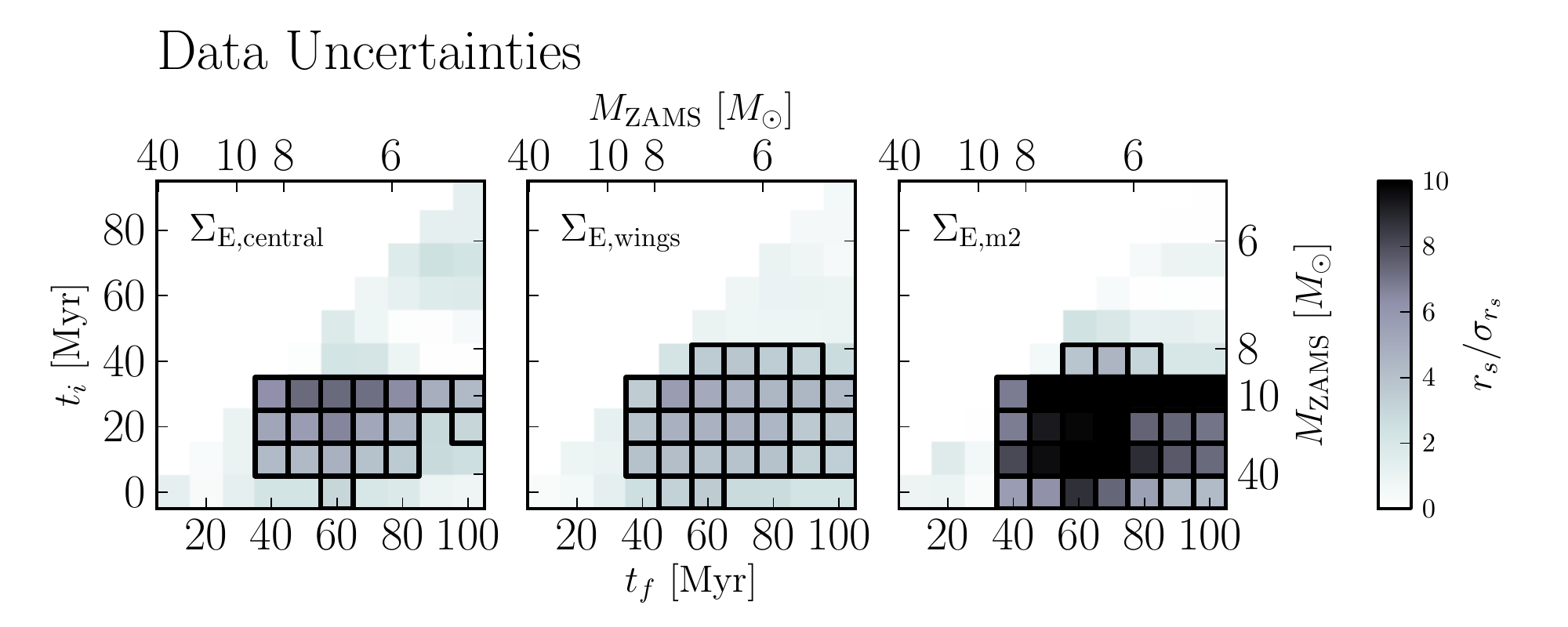}
\caption[Correlation coefficients between \sfrsd{} and $\Sigma_\mathrm{E, HI}$]{
Correlation coefficients \rs{} and associated uncertainties between \avesfrsd{t_i \rightarrow t_f} and \hi{} energy surface densities for $\Sigma_\mathrm{E, central}$, $\Sigma_\mathrm{E, wings}$, and $\Sigma_\mathrm{E, m2}$ for the entire sample.
The upper panel shows the \rs{} values themselves, with the $x$-axis representing $t_i$ values and the $y$-axis representing $t_f$ values.
We have shown the masses of stars whose ages correspond to the $t_i$ and $t_f$ values ($\m{ZAMS}$) on the upper and right axes.
The color of a single box represents the \rs{} value between that \hi{} energy surface density and \avesfrsd{t_i \rightarrow t_f} for the entire sample.
Blue indicates a positive correlation ($\rs{} > 0$), red indicates a negative correlation ($\rs{} < 0$), and white indicates no correlation ($\rs{} = 0$).
The lower two rows show the significance ($\rs{} / \sigma_{r_s}$) of the correlation coefficients in the upper panel due to bootstrapping and data uncertainties, with darker colors indicating more significant correlations.
We have outlined boxes with $\rs{} / \sigma_{r_s} > 3$ in black.
We find significant correlations between \hi{} energy surface density and \avesfrsd{t_i \rightarrow t_f} on timescales of $\sim 30 - 60$ Myr.
\label{angst::fig:corr-ti-tf}
}
\end{figure}

In the upper row of Figure~\ref{angst::fig:corr-ti-tf}, we present the correlation coefficients between \avesfrsd{t_i \rightarrow t_f} and $\Sigma_\mathrm{E,central}$ (right panels), $\Sigma_\mathrm{E,wings}$ (middle panels), and $\Sigma_\mathrm{E, m2}$ (left panels).
As in Figure 5, each bin represents a specific time interval defined by $t_i$ and $t_f$, and the color coding indicates the sign and strength of the correlation.
The lower two rows of Figure~\ref{angst::fig:corr-ti-tf} show the significance of the correlation coefficients in the upper row, based on bootstrapping and data uncertainties.
In general, the bootstrap uncertainties provide more stringent limits on the significance of measured \rs{} values.

We find significant correlations between \hi{} energy surface density and \sfrsd{} over timescales that include SF between $30 - 60$~Myr. The \hi{} energy in the central peak is most strongly correlated with \sfrsd{} at $t \sim 30 - 40$ Myr, while the energy in the wings is most correlated at slightly later times ($t \sim 30 - 60$ Myr).
As expected, the correlation coefficients between \avesfrsd{t_i \rightarrow t_f} and $\Sigma_\mathrm{E, m2}$ are a combination of those with $\Sigma_\mathrm{E,central}$ and $\Sigma_\mathrm{E, wings}$.
We also note that there are no significant correlations when only SF older than 40~Myr is included.

We also find correlations with time ranges that include the $30 - 40$~Myr time bin but cover a longer time range (i.e., the bins to the lower right of the $30 - 40$~Myr bins).
It is likely that these correlations are primarily due to a correlation with the $30 - 40$~Myr bin alone.
To test this idea, we generate random SFHs for each galaxy with the same median SFR as measured within the past 100 Myr and with Gaussian fluctuations based on the standard deviation of the SFR for each galaxy.
We then set the SFR between $30-40$~Myr to its actual value for each galaxy, which allows us to test whether the underlying correlation at $30 - 40$~Myr is responsible for the observed correlations with time ranges that include the $30-40$~Myr interval.
We then calculate the correlation coefficients as for the real data.

\begin{figure}
\centering
\includegraphics[width=3in]{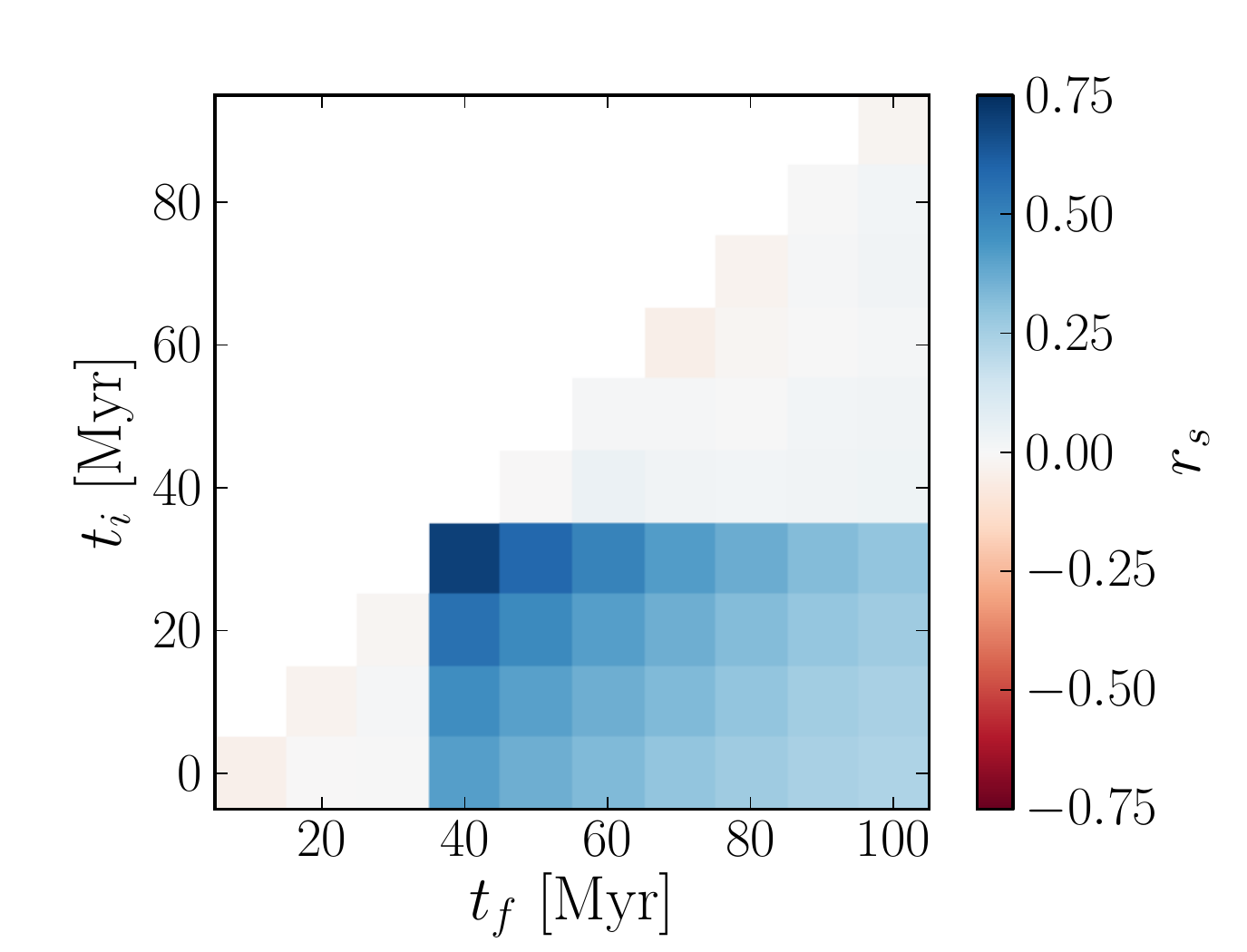}
\caption[Correlation coefficients for randomly-generated SFHs]{ Mean Spearman correlation coefficients between the observed $\Sigma_\mathrm{E,central}$ and SFHs generated randomly except between $30-40$~Myr, where we have set the SFR to the observed SFR.
The sharp edges seen at $t_i < 30$~Myr and $t_f > 40$~Myr are solely an artifact of the correlation with SF between $30-40$~Myr.
\label{angst::fig:corr-ti-tf-random}
}
\end{figure}

In Figure~\ref{angst::fig:corr-ti-tf-random}, we plot the average \rs{} values between \hi{} energy surface density and \avesfrsd{t_i \rightarrow t_f} for 20 randomly-generated SFHs with the imposed correlation at $30-40$ Myr.
The correlation coefficients for bins that include the $30-40$~Myr bin still appear correlated due to the imposed correlation at $30-40$~Myr even though the SFRs in other time ranges were randomly-generated, while time bins that do not include the $30-40$~Myr period are completely uncorrelated.
The sharp edges in Figure~\ref{angst::fig:corr-ti-tf} are most likely due to this effect.
The correlations for $\Sigma_\mathrm{E,wings}$ do not show as sharp edges, suggesting that a wider range of time bins contributes to driving the kinematics of the wings.

\begin{figure}
\centering
\includegraphics[width=6in]{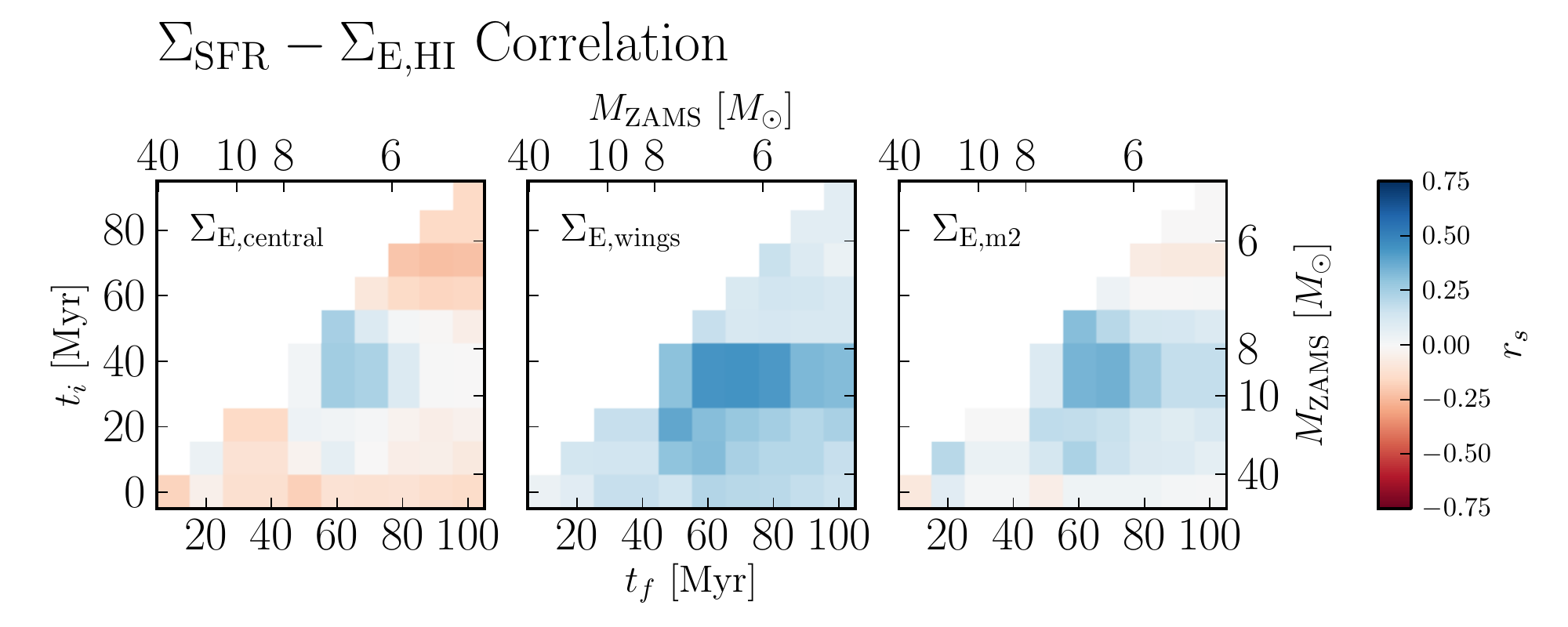}
\caption[Correlation coefficients when ignoring SF between $30-40$~Myr ago]{ The correlation coefficients when SF between 30-40 Myr ago is ignored.
The layout of the figure is the same as the upper panel of Figure~\ref{angst::fig:corr-ti-tf}.
The strong correlations with longer time ranges that include the $30-40$~Myr time range no longer exist.
\label{angst::fig:corr-ti-tf-ignore}}
\end{figure}

To test if there are any correlations that remain when SF between $30-40$~Myr ago is excluded, we artificially set the SFR between $30-40$~Myr to 0 \msun{} \per{yr} \per[2]{kpc} and re-calculate the correlation coefficients.
The results of this test are shown in Figure~\ref{angst::fig:corr-ti-tf-ignore}.
The positive correlations with other time ranges that include the $30-40$~Myr range no longer exist, indicating that the SF between $30-40$~Myr is the main correlation.
For $\Sigma_\mathrm{E,central}$, the correlations are gone entirely.
For $\Sigma_\mathrm{E,wings}$, there are still hints of correlations, also suggesting that the wing kinematics also include input from SF $40 - 60$~Myr ago.

\begin{figure}
\centering
\includegraphics[height=3in]{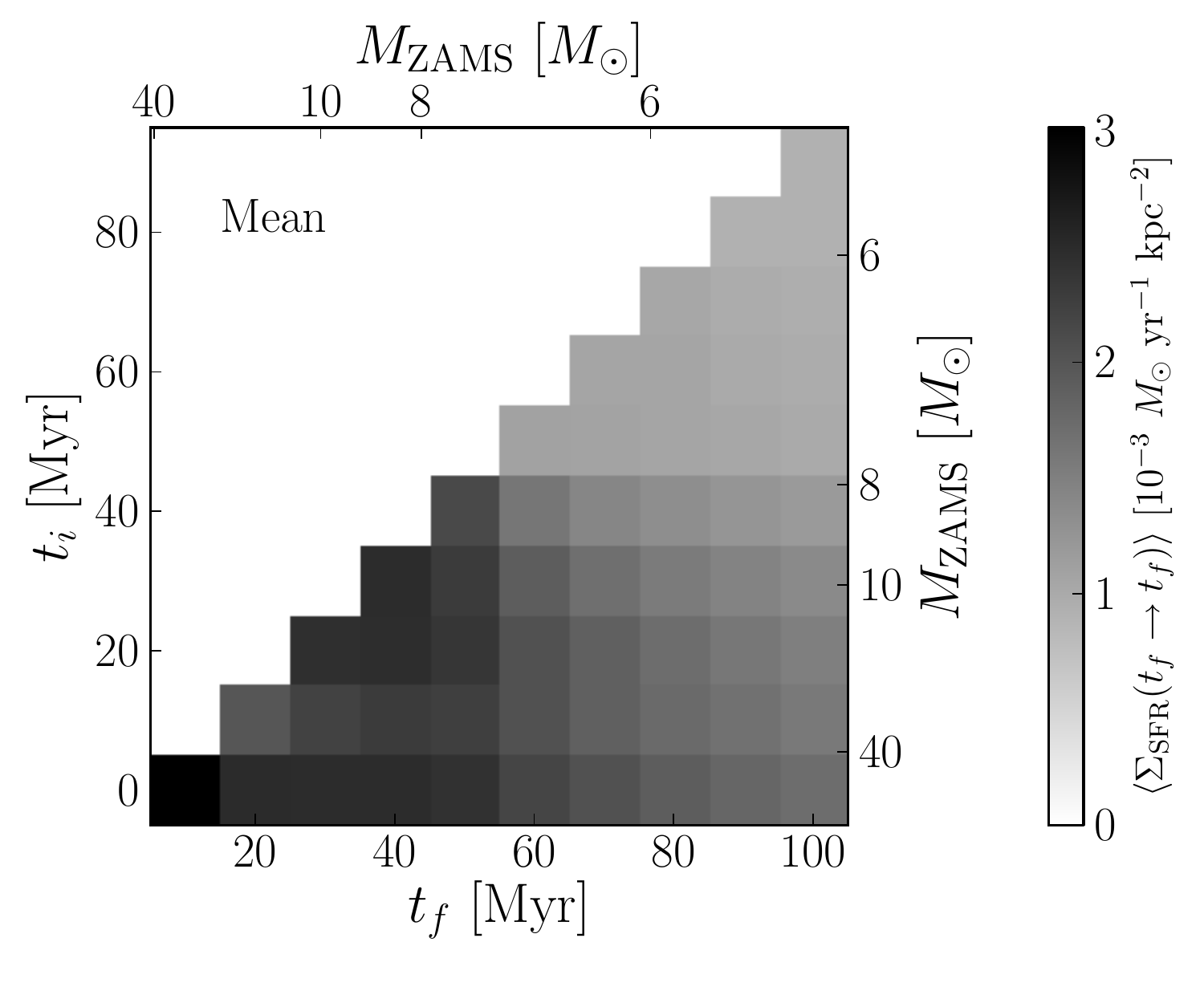}
\includegraphics[height=3in]{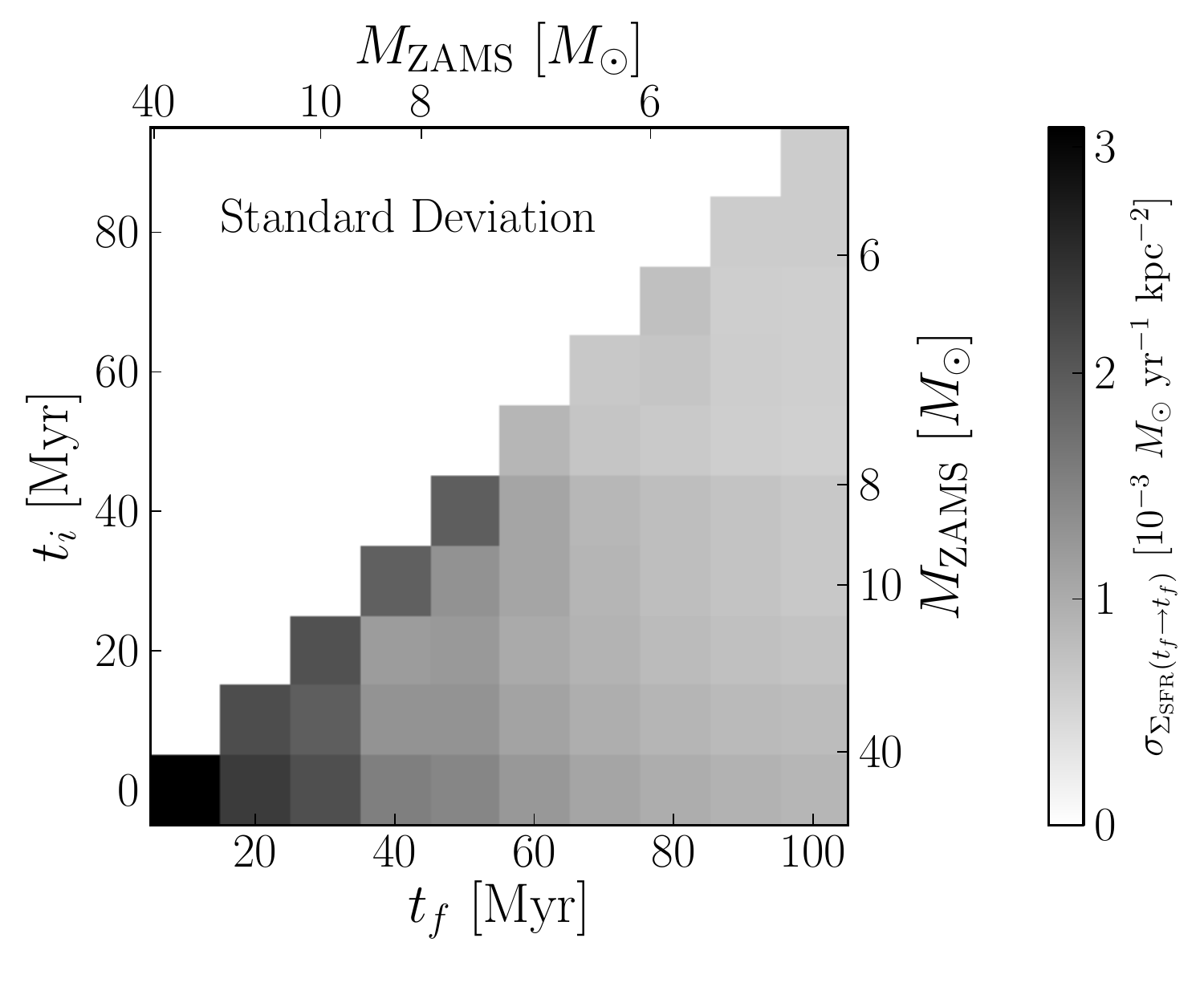}
\caption{ The mean (upper panel) and standard deviation (lower panel) \sfrsd{} of the sample, averaged in 10~Myr bins. The $30-40$~Myr timescale does not show any distinctive features. The increased standard deviation of the diagonal boxes in the lower panel correspond to the shortest sampling of the SFHs, which have lower signal-to-noise.
\label{angst::fig:sfhs-mean-std}}
\end{figure}

Finally, we must explore whether the SFHs of the sample show any characteristic features on any timescale.
In Figure~\ref{angst::fig:sfhs-mean-std}, we plot the mean and standard deviation of the SFHs for the entire sample for the same $t_i \rightarrow t_f$ ranges as in Figure~\ref{angst::fig:corr-ti-tf}.
The mean \sfrsd{} of the sample, shown in the upper panel, is not any stronger at $30 \rightarrow 40$~Myr than surrounding time ranges.
Similarly, as seen in the lower panel, the standard deviation of the SFHs is not highest on the $30 - 40$~Myr timescale.
We note that the diagonal boxes correspond to the shortest sampling of the SFHs, which have the highest errors, and thus increase the standard deviation.
The correlation between \hi{} energy and \avesfrsd{30 \rightarrow 40~\mathrm{Myr}} is therefore unlikely to be representative of a characteristic feature in the SFHs on that timescale.

\begin{figure}
\centering
\includegraphics[width=4in]{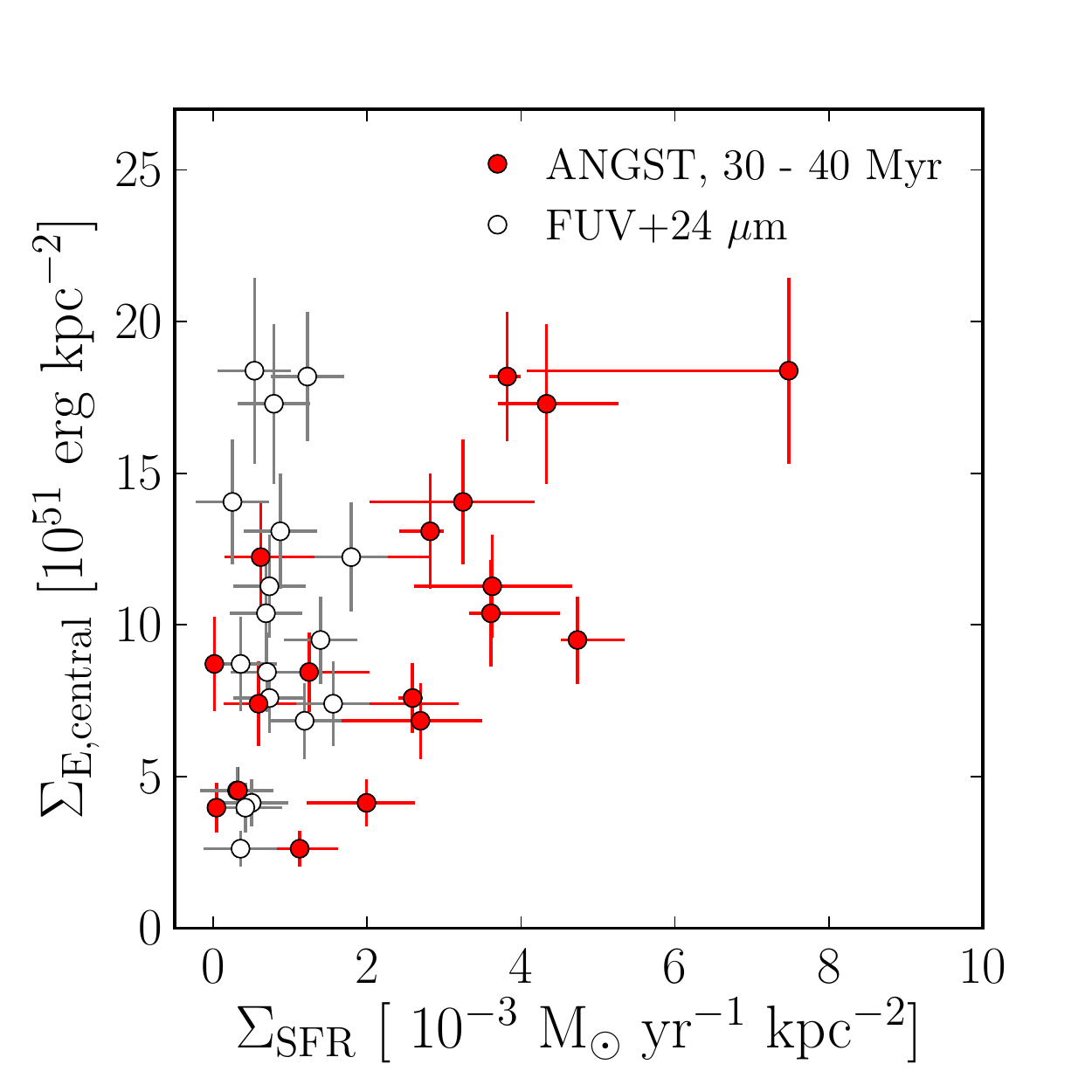}
\caption[]{The correlation between $\Sigma_\mathrm{E, central}$ and both $\avesfrsd{30 \rightarrow 40 \; \mathrm{Myr}}$, measured with ANGST SFHs, and \sfrsd{}, measured with FUV+24\um{}.
The correlation is robust when SF is measured with ANGST SFHs, but disappears when FUV+24\um{} emission is used as a SFR tracer.
\label{angst::fig:sfr-fuv-ti-tf}
}
\end{figure}

To illustrate the improvement that time-resolved SFHs provide, we plot $\Sigma_\mathrm{E, central}$ versus both \avesfrsd{30 \rightarrow 40 \; \mathrm{Myr}}, measured with using ANGST SFHs, and \sfrsd{}, measured with FUV+24\um{} emission in the ANGST aperture in Figure~\ref{angst::fig:sfr-fuv-ti-tf}.
We show \sfrsd{} derived from ANGST SFHs as the filled red circles, and that derived from FUV+24\um{} emission as unfilled black circles.
We have followed the procedure outlined in \citet{StilpGlobal} to calculate FUV+24\um{} SFRs; as with the other measurements, we include only pixels in the ANGST aperture for this calculation.
The \rs{} value for \sfrsd{} derived from ANGST SFHs is $\rs = 0.71$ ($\ps{} = 0.001$), but drops to $0.30$ ($\ps{} = 0.233$) for \sfrsd{} derived from FUV+24\um{} emission.
The underlying correlation with \avesfrsd{t_i \rightarrow t_f} using the ANGST SFHs is when \sfrsd{} is measured with FUV+24\um{}.
We note that the correlation between $\Sigma_\mathrm{E,central}$ and \avesfrsd{30 \rightarrow 40 \; \mathrm{Myr}} becomes even stronger ($\rs{} = 0.8$; $\ps{} < 10^{-3}$) when the galaxies with $\Sigma_\mathrm{SFR} < 10^{-3}$ \msun{} \per{yr} \per[2]{kpc} are excluded.

\section{Discussion}
\label{angst::sec:discussion}

Now that we have assessed the timescales over which \hi{} energy surface density is most strongly correlated with \avesfrsd{t_i \rightarrow t_f}, we can discuss the physical causes behind the correlations.
In \S~\ref{angst::sec:comparisons}, we found that the average turbulent \hi{} energy surface density, traced by $\Sigma_\mathrm{E,central}$, shows a strong correlation with \sfrsd{} at $t \sim 30 - 40$  Myr, suggesting that the neutral ISM is most affected by SF that occurred approximately $30 - 40$ Myr in the past.
The lack of any correlation with SF younger than 30 Myr or older than 40 Myr likewise suggests that younger or older SF has no significant coupling to the ISM.
The energy in the wings is more strongly correlated with SF $\sim 30 - 60$~Myr in the past.
This may suggest a longer coupling timescale for kinematic motions in the wings compared to the central peak, but the difference in correlation coefficients compared to the $30 - 40$~Myr timescale is not large.

We focus primarily on the correlation between the \sfrsd{} and the \hi{} energy surface density of the central peak for the remainder of this section.
In \S~\ref{angst::sec:discussion--timescales}, we explore what physical processes could be responsible for this timescale.
In \S~\ref{angst::sec:discussion--efficiency}, we investigate the implied efficiencies of coupling SF energy to \hi{} energy in light of the observed correlation.

\subsection{SF Processes on the 30-40 Myr Timescale}
\label{angst::sec:discussion--timescales}

The $30 - 40$ Myr timescale is similar to the lifetime of an 8 \msun{} star \citeeg{Girardi1996, Girardi2000, Prialnik2009}, which is approximately the observationally-determined minimum mass of Type II SNe progenitors \citep{Smartt2009, Jennings2012}.
Due to the steepness of the initial mass function (IMF), the majority of SNe progenitors are $\sim 8$ \msun{}. 
These SNe, from lower-mass progenitors, are traditionally thought to release approximately the same amount of mechanical energy into the ISM as their higher mass counterparts.
We therefore might expect that the energy input rate increases with the age of SF burst, as more populous lower-mass progenitors from older SF undergo SNe, up until the time when the stars that can contribute to turbulent energy are too low mass to undergo SNe.
However, this naive assumption is complicated by the fact that the relationship between progenitor mass and lifetime is not linear, such that a wider mass range contributes to a given time bin for high mass stars at recent times.

We can compare these two effects as follows.
We estimate the rate at which energy is input into the ISM due to an instantaneous burst of SF by comparing the IMF to the lifetime of massive stars, following \citet{Shull1995}.
If each SNe emits $E_\mathrm{51} \equiv 10^{51}$ ergs, the energy input rate due to SNe is given by:
\begin{equation}
\frac{d E_\mathrm{SN}}{d t} = E_\mathrm{51} \frac{d}{dt} N_\mathrm{SN} (t),
\label{angst::eqn:de-dt-1}
\end{equation}
where $N_\mathrm{SN} (t)$ is number of supernova as a function of time.
We can represent $d N_\mathrm{SN} / dt$ as $[d N / dm] [dm / dt]$.
The $(dN / dm)$ factor is simply the IMF, which can be written as $(d N / d m) \propto m^{-\alpha}$.
To first order, we can also assume that the lifetime of a star, $(d m / d t)$, is also given by a power law, where $t \propto m^{-\tau}$.
Equation~\ref{angst::eqn:de-dt-1} can therefore be written as:
\begin{equation}
\frac{d E_\mathrm{SN}}{d t} \propto \frac{d}{dt} t^{(\alpha - 1) / \tau} \propto t^{[(\alpha - 1) / \tau] - 1}.
\label{angst::eqn:de-dt-2}
\end{equation}
If we adopt $\alpha = 2.3$ for the upper end of the IMF \citeeg{Kroupa2001} and $\tau = 2.6$ for the mass-lifetime relationship of massive stars \citeeg{Prialnik2009}, we find that $dE / dt \propto t^{-0.5}$, a scaling similar to that found in STARBURST99 \citep[where $dE / dt \propto t^{-0.45}$;][]{Leitherer1999}.
The inverse scaling between $dE / dt$ and $t$ implies that the majority of the energy is released shortly after the burst, contrary to what we might have expected based on only the IMF.
Even though the most massive stars are the least populous, the spread in their ages is very small compared to the lower mass SNe progenitors.
To produce an energy input rate that increases with time, such that $dE / dt$ peaks at later times, we would need to either increase the slope of the IMF ($\alpha$) or decrease the slope of the mass-lifetime relationship for massive stars ($\tau$), such that $(\alpha - 1) / \tau - 1 > 0$.
Literature estimates of $\alpha$ range from $\sim 1.5 - 4$, but many of the measurements also have large uncertainties \citep[][ and references therein]{Weisz2013}.
For $\tau$ to remain the same and still produce $dE / dt$ that increases with time, $\alpha$ must increase to $> 3.6$, a value well outside the weighted mean of $\langle \alpha \rangle = 2.46 \pm 0.35$ for the literature values quoted in \citet{Weisz2013}.
On the other hand, recent results from \citet{Jennings2012} for SN progenitors find that their data are inconsistent with $\alpha$ values outside the range $2.7 - 4.4$.
The \citet{Jennings2012} study also finds fewer massive progenitors than expected, which may could help mitigate this mismatch but is unlikely to fully solve it.
Second, if $\alpha$ were fixed to 2.3, $\tau$ must decease to $< 1.3$, which defies all that is known about stellar evolution.
It is unlikely that we could find $\alpha$ and $\tau$ values that conspire to produce a $dE / dt$ relationship that increases back to $30 - 40$~Myr ago.

If we do not assume that the energy of an individual supernova is fixed, and instead is related mass as $E \propto m^{-\xi}$, we derive a scaling relation given by:
\begin{equation}
\frac{d E_\mathrm{SN}}{d t} \propto t^{ [ (\alpha + \xi - 1) / \tau ] - 1}.
\end{equation}
For the total SNe energy input rate to increase with time, given the above values of $\tau = 2.6$ and $\alpha = 2.3$, we find that $\xi > 2.3$, or $E_\mathrm{SN} \propto m^{-2.3}$, which implies a rather unrealistic scaling.
Moreover, recent research indicates that the supernova energy may increase with increasing mass \citep{Janka2012}, which is the opposite behavior required to produce a peak energy release rate at later times.
%\textbf{Add text about SN energy as a function of mass.}

The scaling relation in Equation~\ref{angst::eqn:de-dt-2} implicitly assumes that the IMF is well-sampled, which only occurs when a large number of stars are formed. 
In the low-mass, low-SFR regime of our sample, however, the high-mass end of the IMF is sampled stochastically, so the statistical approach presented by Equation~\ref{angst::eqn:de-dt-2} does not necessarily apply to our sample.
Many of the traditional methods for estimating luminosity, spectra, and energy input from SF assume a well-sampled IMF and therefore do not account for stochasticity (e.g., STARBURST99, \citet{Leitherer1999}; GALEV, \citet{Kotulla2009}).
Recently, studies have begun to characterize the effects of stochastically sampling the IMF, especially at the low SFRs representative of our sample \citeeg{Lee2009, Weisz2012, daSilva2012}.
The effects of stochasticity on the output SF energy are complicated, especially when considering time-resolved SFHs, and can only be ignored for $\mathrm{SFR} \gg 10^{-2}$ \msun{} \per{yr} \citep{daSilva2012}, but we can estimate its effects to first order by considering the median lifetime of SNe progenitors in stellar clusters of a given total mass.
For each cluster, we draw $N$ stars from an IMF with $\alpha = 2.3$ such that the total mass of the cluster is within 10\% of the desired mass, assuming an instantaneous burst.
We then calculate the median time after the burst that SNe exploded, using the mass-lifetime relationship above (i.e., $t \propto m^{-2.6}$), and then find the median time after the burst for all the SNe in the cluster.

\begin{figure}
\centering
\includegraphics[height=6in]{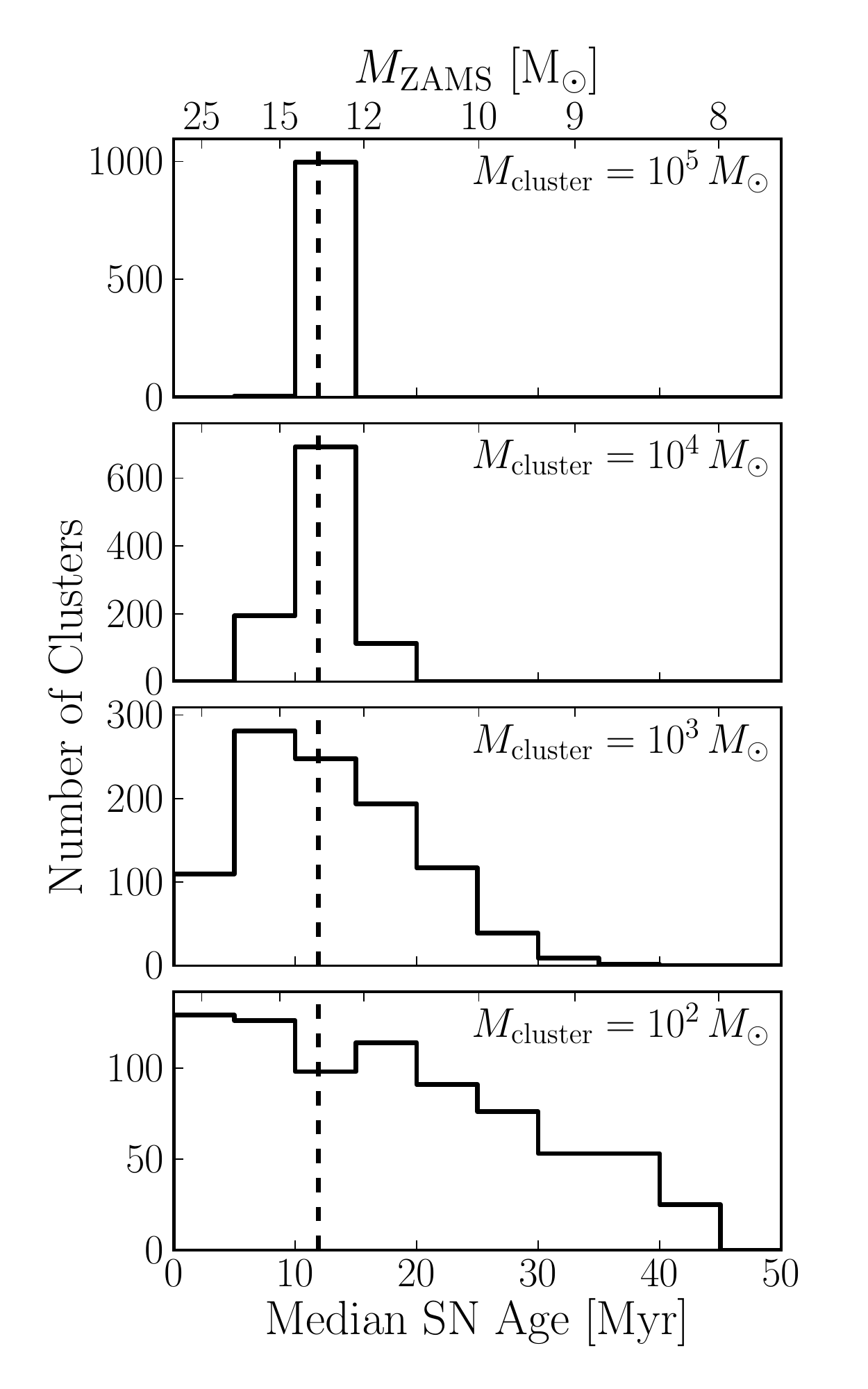}
\caption[Effects of stochasticity on median SN progenitor lifetime]{Effects of stochasticity on the median lifetime of SN progenitors in clusters of different masses.
Each panel shows the histogram of median SN progenitor lifetimes for clusters of the same mass, with mass decreasing in lower panels.
The dashed vertical line represents the expected median SN progenitor lifetime for a well-sampled IMF.
Clusters of $M_\mathrm{cluster} = 10^5$ \msun{} typically approach the expected median age, but stochasticity becomes increasingly important for clusters of $M_\mathrm{cluster} \sim 10^{2} - 10^{3}$ \msun{}.
We have overlaid the approximate $M_\mathrm{ZAMS}$ corresponding to the median SN progenitor lifetime, assuming that $t \propto m^{-2.6}$.
\label{angst::fig:stochasticity}
}
\end{figure}

In Figure~\ref{angst::fig:stochasticity}, we plot the histograms of the median age of SNe for 100 multiple realizations of clusters of a given mass.
The dashed vertical line in each panel represents the expected median SN progenitor lifetime for a well-sampled IMF.
For clusters of $M_\mathrm{cluster} = 10^5$ \msun{}, the lifetime of the median SN progenitor approaches the expected value for a well-sampled IMF.
At lower total masses (i.e., $M_\mathrm{cluster} = 10^2 - 10^3$ \msun{}), the median SN progenitor lifetime is not well-defined and depends on random sampling of the IMF.
We note that this is a simplistic model for stochasticity in these galaxies, but it provides a back-of-the-envelope estimation of whether stochasticity is important for the sample.

We can compare these masses to the average mass in stars formed by the sample galaxies over a 10~Myr time period.
The lowest-mass galaxies have an average SFR over the past 100~Myr of $\sim 10^{-3}$ \msun{} \per{yr}, implying that they form on average $\sim 10^4$ \msun{} over 10 Myr.
If all the SF over the 10 Myr interval were concentrated in one burst, Figure~\ref{angst::fig:stochasticity} indicates that stochasticity is not important even for the lowest-mass dwarfs.
However, it is more likely that SF has been occurred in individual clusters, each with a lower mass.
When the cluster mass function is also taken into account, \citet{daSilva2012} estimate that stochastic effects are negligible only for SFR $\gg 10^{-2}$ \msun{} \per{yr}, a condition that is not met by any galaxy in our sample.

The observed correlation with the $30 - 40$~Myr timescale is the time between when stars form and when those stars leave an observable signature in the turbulence of the neutral ISM.
However, that cycle consists of several steps: the timescale for SF to inject energy into the local environment, and timescale for that local energy input to be transferred into the larger reserve of the neutral medium.
We now estimate each of these timescales.
The first timescale describes the timescale for the peak energy release rate from SF, assuming a well-sampled IMF.
However, this energy is released in powerful SNe, which shock heat the surrounding ISM, and may not be immediately observable as \hi{} turbulence.
The second timescale is related to the propagation of supernova remnants (SNR) through the ISM.
A lower limit for the second timescale is the time for an individual SNR to merge with the ISM, approximately $\sim 5$ Myr \citep{Cioffi1988}.
Similar behavior is also seen in the semi-analytic models by \citet{Braun2012b}, where the ISM cools to pre-SF conditions $\sim 5$~Myr after the last SN progenitor. 
The back-of-the-envelope relationship between SF and \hi{} energy should therefore be only $\sim 15$ Myr for SF regions that adequately sample the high-mass end of the ISM.
On the other hand, X-ray observations of dwarf starburst galaxies indicate that the cooling timescale of hot gas ranges between  $\sim 10 - 200$~Myr, though the results depend on the filling factor of this phase \citep{Ott2005}.
Our results are consistent with SN-driven turbulence if the cooling timescale is $\sim 20 - 30$~Myr in these systems.
However, it is unclear how to translate the simulations of individual SNRs or feedback in the case of a well-sampled IMF to the stochastic SF in our sample.

Because the timescale for peak energy input from SNe is not necessarily well-matched to the observed $30-40$ Myr correlation, we also briefly consider the idea that \hi{} line widths are powered by some other stellar mechanism where peak energy input occurs at later times than SNe.
One source of stellar energy is feedback from high-mass X-ray binaries (HMXBs), which are thought to provide $\sim 0.1-1$ times the energy from SNe and could therefore contribute if the coupling efficiency is higher than that for SNe \citep{Justham2012}.
Recent observations of HMXBs find that the majority have ages of $40-60$~Myr, implying that they are most populous $\sim 40$~Myr after a star formation event \citep{Antoniou2010, Williams2013}.
Other proposed stellar sources of energy are stellar winds \citeeg{Abbott1982, VanBuren1985} and ionizing radiation \citeeg{Kritsuk2002, Kritsuk2002a}.
In dwarfs, simulations by \citet{Hopkins2012} show that radiation pressure from stellar winds has very little effect compared to SNe on the surrounding ISM due to low gas densities and metallicities.
\citet{MacLow2004} also estimate the energy input due to stellar winds, and suggest that a substantial energy input is seen only from the most massive Wolf-Rayet stars, which have a much shorter lifetime than the timescale for peak energy input from SNe.
In addition, winds from AGB stars typically have small wind velocities \cite[$\ll 100$ \kms{}; e.g.,][]{Knapp1985, Loup1993} and therefore cannot impart as much energy to the ISM \citeeg{Oppenheimer2008}.
Similarly, ionizing radiation from stars is not expected to be a large source of energy for turbulence \citeeg{MacLow2004}.
In addition, the ionizing radiation is stronger from the most massive stars, again with the shortest lifetimes, and thus does not explain the observed $30-40$~Myr timescale.
The other mechanisms for stellar energy input therefore seem unlikely as an explanation for the observed timescale.

\subsection{Coupling Efficiency between SF Energy and \hi{} Energy}
\label{angst::sec:discussion--efficiency}

Recent observations have shown that SF cannot provide enough energy to account for the observed energy in \hi{} at low \sfrsd{} \citeeg{Tamburro2009, StilpResolved}.
However, these studies used FUV+24\um{} emission to measure SFRs in the sample galaxies, which misses the correlation between $\Sigma_\mathrm{E,central}$ and $\Sigma_\mathrm{SFR}$ as measured with ANGST on $30 - 40$ Myr timescales (see Figure~\ref{angst::fig:sfr-fuv-ti-tf}).
We now measure whether the better measurement of SFR can fix the problem of unrealistic efficiencies in the low \sfrsd{} regime.

We estimate the energy input from SNe over the $30-40$ Myr timescale as follows:
\begin{equation}
\Sigma_\mathrm{E,SF} = \eta \; \left[\avesfrsd{30 - 40 \; \mathrm{Myr}} \times 10 \; \mathrm{Myr} \right] \; E_\mathrm{51},
\label{angst::eqn:e_sf}
\end{equation}
where $\eta$ is the number of SN per unit solar mass formed.
The quantity $\eta (\Sigma_\mathrm{SFR} \times 10 \; \mathrm{Myr})$ represents the total number of SNe per area due to SF over the $30-40$ Myr timescale.
We set $\eta = 3.3 \times 10^{-3}$ to be the fraction of stars with $M > 8$ \msun{} for a single-slope power law ($\alpha = -2.3$) with mass limits of 0.1 \msun{} and 120 \msun{}.
This equation gives us the total energy surface density from SNe due to SF that occurred over the past $30 - 40$~Myr.
We note that we have not included energy input due to SF that formed at more recent times.
Our estimate of $E_\mathrm{SF}$ is therefore a lower limit, with the caveat that the fiducial dissipation timescale for turbulence is expected to be $\sim 10$~Myr \citep{MacLow1999}.
It is possible that SNe from stars that formed more recently than 30~Myr ago have contributed to the \hi{} turbulent energy.
If that were the case, however, we might have expected to see a stronger correlation with time ranges that also included more recent times.

\begin{figure}
\centering
\includegraphics{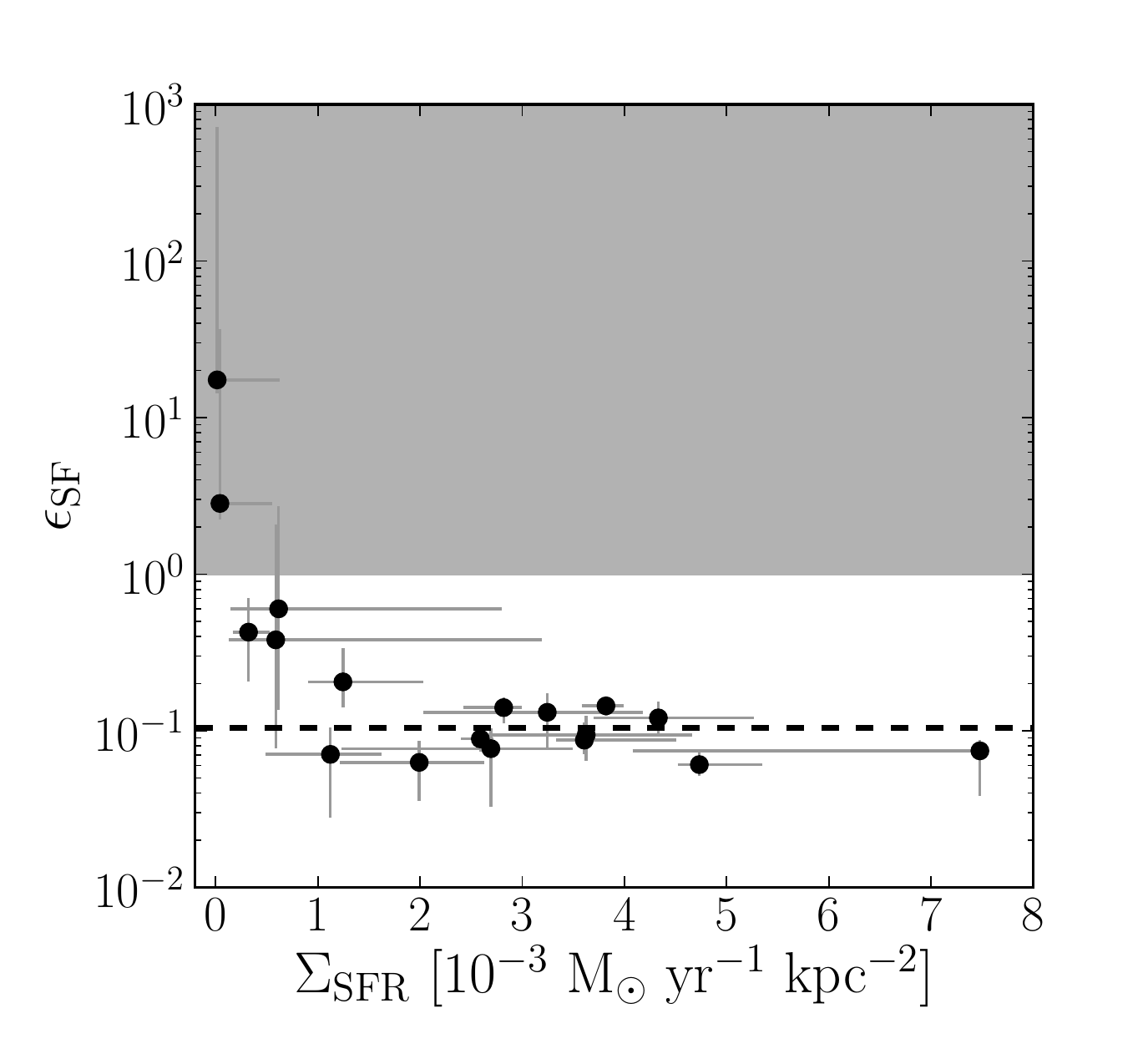}
\caption[Implied coupling efficiency between SF energy and \hi{} energy.]{The implied coupling efficiency between SF energy and \hi{} energy, where $\epsilon_\mathrm{SF} \equiv \Sigma_\mathrm{E,HI} / \Sigma_\mathrm{E,SF}$.
At $\sfrsd{} < 10^{-3}$ \msun{} \per{yr} \per[2]{kpc}, the efficiencies approach and can exceed unphysical values of $\epsilon_\mathrm{SF} > 1$.
At higher \sfrsd{}, we find constant efficiencies, with a mean and standard deviation of $\epsilon_\mathrm{SF} = 0.10 \pm 0.04$ (shown as the dashed line).
\label{angst::fig:efficiency}
}
\end{figure}

In Figure~\ref{angst::fig:efficiency}, we plot the efficiency required to explain the observed \hi{} energy with the energy input from SF.
We define $\epsilon_\mathrm{SF} \equiv \Sigma_\mathrm{E,HI} / \Sigma_\mathrm{E,SF}$.
At $\sfrsd{} < 10^{-3}$ \msun{} \per{yr} \per[2]{kpc}, we find that efficiencies of $> 0.1$ are required to couple SF energy to \hi{} energy.
At higher \sfrsd{}, however, we find that the efficiency approaches a constant value of $0.11$ with a standard deviation of $0.04$ across our sample.
The efficiencies at $\sfrsd{} > 10^{-3}$ \msun{} \per{yr} \per[2]{kpc} are also similar to those estimated from simulations \citep[$\sim 0.1$;][]{Thornton1998, Krause2013}, though other simulations have estimated higher efficiencies \citep[e.g., 0.5;][]{TenorioTagle1991}.
The $\sfrsd{} = 10^{-3}$ \msun{} \per{yr} \per[2]{kpc} limit is approximately the threshold where the relationship between SF energy and \hi{} energy has been seen to break down \citeeg{Dib2006, Tamburro2009, StilpResolved}, so our results are consistent with these studies.

The high efficiencies required at low \sfrsd{} can exceed theoretical limitations, and, at the lowest \sfrsd{}, are higher than the unphysical limit of $\epsilon_\mathrm{SF} > 1$ where \hi{} has more energy than SF provides.
As discussed in a forthcoming paper \citep{StilpResolved}, this behavior implies that the \hi{} line widths in the lowest \sfrsd{} regime are driven by processes other than SF even when SF is properly matched in timescale.
Such processes could provide a base level of \hi{} turbulent velocity dispersions in all regions of the ISM, which is then enhanced by the presence of star formation.
However, the majority of proposed non-SF turbulence drivers require shearing motions \citeeg{Kim2007, Agertz2009} and are therefore likely to be less effective in dwarfs, which shear is weaker.

The above estimates of efficiency assume a well-sampled IMF.
The galaxies with the largest \avesfrsd{30 - 40 \; \mathrm{Myr}} form $10^5$ \msun{} stars over that time range, with those with the smallest only form $\sim 250$ \msun{} of stars.
Stochasticity may therefore be important for the galaxies with the smallest SFRs, as clusters of smaller mass are more likely to form more low mass stars instead of a few high mass stars compared to clusters of larger total mass.
However, the simulated clusters discussed in \S~\ref{angst::sec:discussion--timescales} with $\m{total} \sim 250$ \msun{} have only approximately $0 - 7$ stars with $\m{} > 8$ \msun{}, with a peak at 2, compared to the expected $0.8$ from Equation~\ref{angst::eqn:e_sf}.
To reach an efficiency similar to that observed at higher \sfrsd{}, a larger number of SNe would be required than what we would expect from stochastic effects.
The unphysical efficiencies at the lowest \sfrsd{} values are therefore not likely to be remedied by stochasticity.

\section{Summary}
\label{angst::sec:summary}

We have compared the energy in \hi{} to the SF surface density, averaged over different timescales, in a number of nearby dwarf galaxies.
We find that the \hi{} energy surface density is correlated with SF that occurred $30 - 40$ Myr ago and shows no correlations at times that do not include this range.
These correlations are washed out when the broadband FUV+24\um{} tracer is used to measure SFR.
This timescale is similar to the theoretical dissipation timescale, but delayed by a time approximately equal to the lifetime of the lowest-mass supernova progenitor, supporting the idea that SNe are a contributing factor to turbulence in the neutral ISM.
However, the stochastic nature of SF in the low SFR regime of our sample galaxies complicates a straightforward explanation.
Second, we find that a constant coupling efficiency of $0.11 \pm 0.04$ galaxies with average $\sfrsd{} > 10^{-3}$ \msun{} \per{yr} \per[2]{kpc} between $30 - 40$ Myr ago, while galaxies with lower \sfrsd{} values require higher or unphysical efficiencies.

We note that we have averaged SF and \hi{} energy properties over the entire ANGST aperture, which in some cases can cover a large area of the galactic disk with widely-varying SF properties.
Similar studies that include spatially-resolved as well as time-resolved SF properties can place more stringent constraints on the timescale over which \hi{} energy is affected by SF.

\begin{acknowledgments}
We thank the anonymous referee for helpful suggestions that improved the quality of this paper.
Support for this work was provided by the National Science Foundation collaborative research grant ``Star Formation, Feedback, and the ISM: Time Resolved Constraints from a Large VLA Survey of Nearby Galaxies,'' grant number AST-0807710.  % NSF grant
This material is based on work supported by the National Science Foundation under grant No. DGE-0718124 as awarded to A.M.S.  %GRFP
The National Radio Astronomy Observatory is a facility of the National Science foundation operated under cooperative agreement by Associated Universities, Inc.  % NRAO
This work is based on observations made with the NASA/ESA Hubble Space Telescope, obtained from the data archive at the Space Telescope Science Institute. Support for this work was provided by NASA through grants GO-10915, DD-11307, and GO-11986 from the Space Telescope Science Institute, which is operated by AURA, Inc., under NASA contract NAS5-26555. %HST
\end{acknowledgments}

\bibliographystyle{apj}
\bibliography{angst}

\end{document}